\documentclass[10pt]{article}

\usepackage[a4paper, margin=1in]{geometry}

\usepackage[authoryear,longnamesfirst]{natbib}
\usepackage{hyperref}

\usepackage{color}
\usepackage{url}
\usepackage[utf8]{inputenc}
\usepackage[T1]{fontenc}
\usepackage{xspace}
\usepackage{xcolor}

\usepackage{paralist}
\usepackage{amssymb}
\usepackage{subfigure}
\usepackage{amsmath}
\usepackage{graphicx}

\usepackage{colortbl}

\usepackage{amsthm}
\usepackage{algorithm}
\usepackage{algpseudocode}
\usepackage{multirow}
\usepackage{xfrac}

\newcommand{\ie}{\emph{i.e.,}\xspace}
\newcommand{\eg}{\emph{e.g.,}\xspace}

\newcommand{\aedes}{\emph{Aedes aegypti}\xspace}

\definecolor{lightgray}{rgb}{0.7, 0.7, 0.7}
\definecolor{vlgray}{rgb}{0.9, 0.9, 0.9}

\begin{document}
\let\WriteBookmarks\relax
\def\floatpagepagefraction{1}
\def\textpagefraction{.001}
\title{Detecting \aedes Mosquitoes through Audio Classification with Convolutional Neural Networks\footnote{Preprint. This work has been submitted to Elsevier for possible publication.}}              
\author{Marcelo Schreiber Fernandes, Weverton Cordeiro, Mariana Recamonde-Mendoza\\
\\
\normalsize  Institute of Informatics (INF)\\
\normalsize  Universidade Federal do Rio Grande do Sul (UFRGS)\\
\normalsize Porto Alegre, Brazil\\
\\
\normalsize marcelo.schreiber@gmail.com, \{weverton.cordeiro,mrmendoza\}@inf.ufrgs.br}
\date{}

\maketitle


\begin{abstract}
The incidence of mosquito-borne diseases is significant in under-developed regions, mostly due to the lack of resources to implement aggressive control measurements against mosquito proliferation. A potential strategy to raise community awareness regarding mosquito proliferation is building a live map of mosquito incidences using smartphone apps and crowdsourcing. In this paper, we explore the possibility of identifying \aedes mosquitoes using machine learning techniques and audio analysis captured from commercially available smartphones. In summary, we downsampled \aedes wingbeat recordings and used them to train a convolutional neural network (CNN) through supervised learning. As a feature, we used the recording spectrogram to represent the mosquito wingbeat frequency over time visually. We trained and compared three classifiers: a binary, a multiclass, and an ensemble of binary classifiers. In our evaluation, the binary and ensemble models achieved accuracy of 97.65\% (± 0.55) and 94.56\% (± 0.77), respectively, whereas the multiclass had an accuracy of 78.12\% (± 2.09). The best sensitivity was observed in the ensemble approach (96.82\% ± 1.62), followed by the multiclass for the particular case of \aedes (90.23\% ± 3.83) and the binary (88.49\% ± 6.68). The binary classifier and the multiclass classifier presented the best balance between precision and recall, with F1-measure close to 90\%. Although the ensemble classifier achieved the lowest precision, thus impairing its F1-measure (79.95\% ± 2.13), it was the most powerful classifier to detect \aedes in our dataset.
\end{abstract}


\section{Introduction}

Although a relevant health issue, mosquito-borne diseases are regarded as largely neglected according to a report from the \cite{worldhealthorganization_2016}. Recent estimates account that nearly 700 million people contract a mosquito-borne illness yearly, with over 1 million fatalities in the same period \citep{caraballo2014emergency}. In 2012, dengue was classified as the viral disease with the highest epidemic potential, with its primary vector being \aedes, which is also a vector for yellow fever, Chikungunya, and Zika virus \citep{world2014global}.
 
A growing concern in the scientific community is that mosquitoes that carry these diseases are becoming resistant to insecticides \citep{who-2020}. Therefore, a potentially more effective form of counteracting the spread of such diseases is limiting mosquito proliferation. In this context, several approaches have been tested to tackle the proliferation of \aedes, for example, identifying places with still water that serve for mosquito reproduction, installing mosquito traps, and releasing genetically modified mos\-quitoes that are unable to reproduce. In parallel, to guide more effective government actions towards limiting the spread of mosquitoes, authorities have attempted to map geographic locations that are more favorable to mosquito incidence and areas in which the mosquitoes have been identified. Despite the efforts, the technical complexity of deploying measurements (for example, installing mosquito traps), associated costs, and required logistics \citep{townson2005exploiting} have made mosquito tracking a largely neglected option.

The lack of situational awareness of mosquito incidence in a region may also contribute to the local community's poor engagement in fighting mosquito breeding sites. One strategy to raise community awareness regarding local mosquito proliferation is building a live map of mosquito incidences using crowdsourcing. For being effective, it should be automated, mainly for avoiding inaccurate reports and minimizing the effort for collecting mosquito incidence data. However, the scarcity of mapping data for these mosquitoes might be caused by the difficulty of acquiring them. In under-developed regions, the identification process involves manual capture and identification. Therefore, it needs human and time resources. To tackle the vectors of mosquito-borne diseases effectively, it is necessary to have complete and updated data on their mapping. Thus, surveillance methods are needed, which are low cost, capable of identifying mosquito populations \citep{mukundarajan2017using}.

Many studies argue that machine learning (ML) algorithms based on supervised learning are the best choice for performing automatic classification of species. However, these algorithms require a high amount of data for training and demand significant computational processing \citep{acevedo2009automated}. In this form of learning, we provide a set of labeled input data to the algorithm, and the algorithm seeks to find a function that is capable of mapping the input data set to their respective labels.

In this paper, we explore the possibility of identifying \aedes mosquitoes using ML techniques. Our identification strategy is analyzing the sound of a wingbeat (recorded, for example, using a smartphone mic) and comparing it with a database of mosquito wingbeat recordings. To this end, we resampled the \aedes wingbeat recordings to 8kHz and used them to train a convolutional neural network (CNN) through supervised learning. 
As a feature, we used the recording spectrogram, so that the mosquito wingbeat frequency is represented visually over time. Three classifiers were trained using distinct approaches: 
(i) binary classifier (\aedes or not); (ii) multiclass classifier (classification between several types of mosquitoes available in the database); and (iii) an ensemble composed of binary classifiers (where each binary classifier performs the classification between \aedes vs. a specific species). An evaluation of the performance of the classifiers has shown the potential of using ML for mosquito identification based on audio analysis\footnote{The datasets of our study can be found at \url{https://github.com/ComputerNetworks-UFRGS/CNN_Aedes_Aegypti_Classifier}}. The sensitivity for detecting \aedes varied between 88.49\% (binary classifier) and 98.82\% (ensemble classifier) and an overall accuracy as high as 97\% was observed in the binary approach. 


The remainder of this paper is structured as follows. In Section~\ref{sec:relwork}, we describe the theoretical framework upon which we built our work, focusing on ML techniques for mosquito wingbeat sound analysis. In Section~\ref{sec:materials-methods}, we review related work, focusing on initiatives to limit the spread of \aedes and mosquito wingbeat analysis. In Section~\ref{sec:approach}, we present the methodology used for \aedes wingbeat analysis, followed by an assessment of its effectiveness in Section~\ref{sec:experiments}. We close the paper in Section~\ref{sec:conclusions} with concluding remarks and perspectives for future research.

\section{Related Work}
\label{sec:relwork}



Strategies for automated identification of mosquitoes using audio features like wingbeat frequency have been investigated as early as 1945 \citep{kahn1945recording} (see \cite{chen2014flying} for a brief historical overview of proposed solutions). Despite the advances achieved, 
\cite{chen2014flying} noted in 2014 that insect classification from audio features had not yet led to inexpensive and ubiquitous solutions, mostly because of the difficulty in obtaining high-quality audio recording samples of mosquito wingbeat. In this context, the research community had been keen to investigate alternate strategies to identify mosquitoes. One direction pursued was using optical sensors like infrared recording device for profiling the mosquito wingbeat and using artificial neural networks to analyze specific recording features \citep{moore1991artificial,li2005automated,chen2014flying,silva2015exploring,potamitis2015insect,ouyang2015mosquito,potamitis2016measuring}. 
\cite{li2005automated}, for example, used a neural network to analyze and identify recorded wingbeat waveforms for five species of mosquitoes, including \aedes. The authors obtained wingbeat waveforms with pseudo-acoustic optical sensors, with mosquitoes placed inside a photosensor connected to the microphone input port of a 16-bit sound card. Other approaches for tracking mosquitoes have also been investigated, for example, image tracking and multiple object tracking. Please refer to \cite{spitzen2018keeping} for a brief survey. More recently, \cite{10.1371/journal.pone.0210829} used CNNs to extract features from mosquito images to identify adult \aedes mosquitoes (as well as \emph{Aedes albopictus} and \emph{Culex quinquefasciatus}), with an accuracy of up to 82\% for female \aedes. 

Despite the enthusiasm with photo sensor-based and image classification based approaches for mosquito identification, exploiting mosquito wingbeat remained a popular approach for building mosquito traps. 
\cite{johnson2016siren} used female flight tones to enhance male collections in non-mechanical passive Gravid Aedes Traps (GAT). 
\cite{balestrino2016sound} also followed a similar approach and identified that \emph{Aedes albopictus} mosquitoes are more responsive to acoustic stimulation up to 4 days of age, and that the black color of the trap influence positively in the effectiveness of the trap. In related work, 
\cite{silva2013applying} enhanced the mosquito trap's capabilities by applying machine learning and audio analysis techniques to identify mosquitoes captured using such traps automatically.

In this paper, we follow the vision that an effective solution for tackling mosquito-borne diseases requires broader engagement from local communities. The use of a ubiquitous device like a smartphone is an appropriate option to aid in such endeavor \citep{wood2019taking}, both because of pervasiveness and capabilities. 
\cite{mukundarajan2017using} found that it is possible to make high-quality mosquito wingbeat recordings using ordinary smartphones, even in scenarios with significant background noise. They also found that the Doppler effect is negligible for such recording since mosquitoes hardly fly at more than one meter per second, with a recording error of around 2 Hz for mosquitoes up to 10cm distant from the microphone. The wingbeat frequency of a mosquito in-flight changes at most in 100Hz, although overlapping in frequencies between mosquitoes of different species is known to occur \citep{brogdon1994measurement}. They also note that smartphone recordings benefit from the fact that microphones are designed to have maximum sensitivity in the frequency range between 300 to 3000Hz (human voice), similar to with the wingbeat frequency of mosquitoes.



Recently, 
\cite{kiskin2020bioacoustic} suggested that bio-acoustic classification using a CNN classifier is a feasible approach for detecting \emph{Culex quinquefasciatus} mosquitoes, with an accuracy of over than 90\%, even with limited training. 
\cite{vasconcelos2019locomobis} also used mosquito recording to identify the occurrence of \aedes in a given area. The authors apply fast Fourier transform (FFT) and quadratic interpolation to extract the dominant frequency in the sample, and compares the result against a small set of known species.  
\cite{staunton2019novel}, in turn, collected wingbeat frequency data by measuring the Doppler shift of a reflected ultrasonic continuous wave signal. These investigations largely benefit from extensive datasets of mosquito wingbeat recording \citep{kiskin2020humbug}, which often provide enough data for training artificial neural networks, or provide distinct audio features one can look for to identify mosquitoes.

\section{Materials and Methods}
\label{sec:materials-methods}
\subsection{Data Collection and Pre-processing} 

\begin{table*}[t]
\centering
\resizebox{0.9\textwidth}{!}{  
\begin{tabular}{lccc}
\hline\hline
Species                                                       & Number of files          & Original duration (seconds) & Duration after pre-processing (seconds) \\ \hline\hline
\aedes                & {22}  & {1,736.87}     & {1,736.87}                 \\ \hline
\emph{Aedes albopictus}             & {7}   & {966.37}       & {966.37}                   \\ \hline
\emph{Aedes mediovittatus}          & {3}   & {180.00}       & {53.69}                    \\ \hline
\emph{Aedes sierrensis}             & {361} & {3,337.58}     & {274.01}                   \\ \hline
\emph{Anopheles albimanus}          & {40}  & {1,736.55}     & {901.37}                   \\ \hline
\emph{Anopheles arabiensis dongola} & {6}   & {1,040.11}     & {850.88}                   \\ \hline
\emph{Anopheles arabiensis rufisque}& {7}   & {943.76}       & {844.45}                   \\ \hline
\emph{Anopheles atroparvus}         & {7}   & {907.64}       & {833.66}                   \\ \hline
\emph{Anopheles dirus}              & {65}  & {1,733.48}     & {530.35}                   \\ \hline
\emph{Anopheles farauti}            & {47}  & {1,727.52}     & {781.35}                   \\ \hline
\emph{Anopheles freeborni}          & {54}  & {2,397.82}     & {1,237.00}                 \\ \hline
\emph{Anopheles gambiae akron}      & {7}   & {972.98}       & {615.15}                   \\ \hline
\emph{Anopheles gambiae kisumu}     & {57}  & {1,747.59}     & {638.57}                   \\ \hline
\emph{Anopheles gambiae rsp}        & {2}   & {567.66}       & {295.84}                   \\ \hline
\emph{Anopheles merus}              & {5}   & {300.00}       & {205.05}                   \\ \hline
\emph{Anopheles minimus}            & {68}  & {2,923.11}     & {994.16}                   \\ \hline
\emph{Anopheles quadriannulatus}    & {7}   & {1,178.10}     & {959.33}                   \\ \hline
\emph{Anopheles quadrimaculatus}    & {6}   & {665.71}       & {548.31}                   \\ \hline
\emph{Anopheles stephensi}          & {58}  & {2,909.69}     & {770.22}                   \\ \hline
\emph{Culex pipiens}                & {9}   & {710.37}       & {240.83}                   \\ \hline
\emph{Culex quinquefasciatus}       & {13}  & {758.13}       & {195.07}                   \\ \hline
\emph{Culex tarsalis}               & {12}  & {613.87}       & {231.11}                   \\ \hline
\emph{Culiseta incidens}                                   & 37                       & 2,971.58                          & 1,244.21                                      \\ \hline\hline
\end{tabular}
}
\caption{Characteristics of the dataset used for model development.}
\label{tab:audiodata}
\end{table*}

\begin{figure}[h]
    \centering
\subfigure[]{%
    \centering
    \label{fig:audio:duration}%
    \includegraphics[width=0.42\columnwidth]{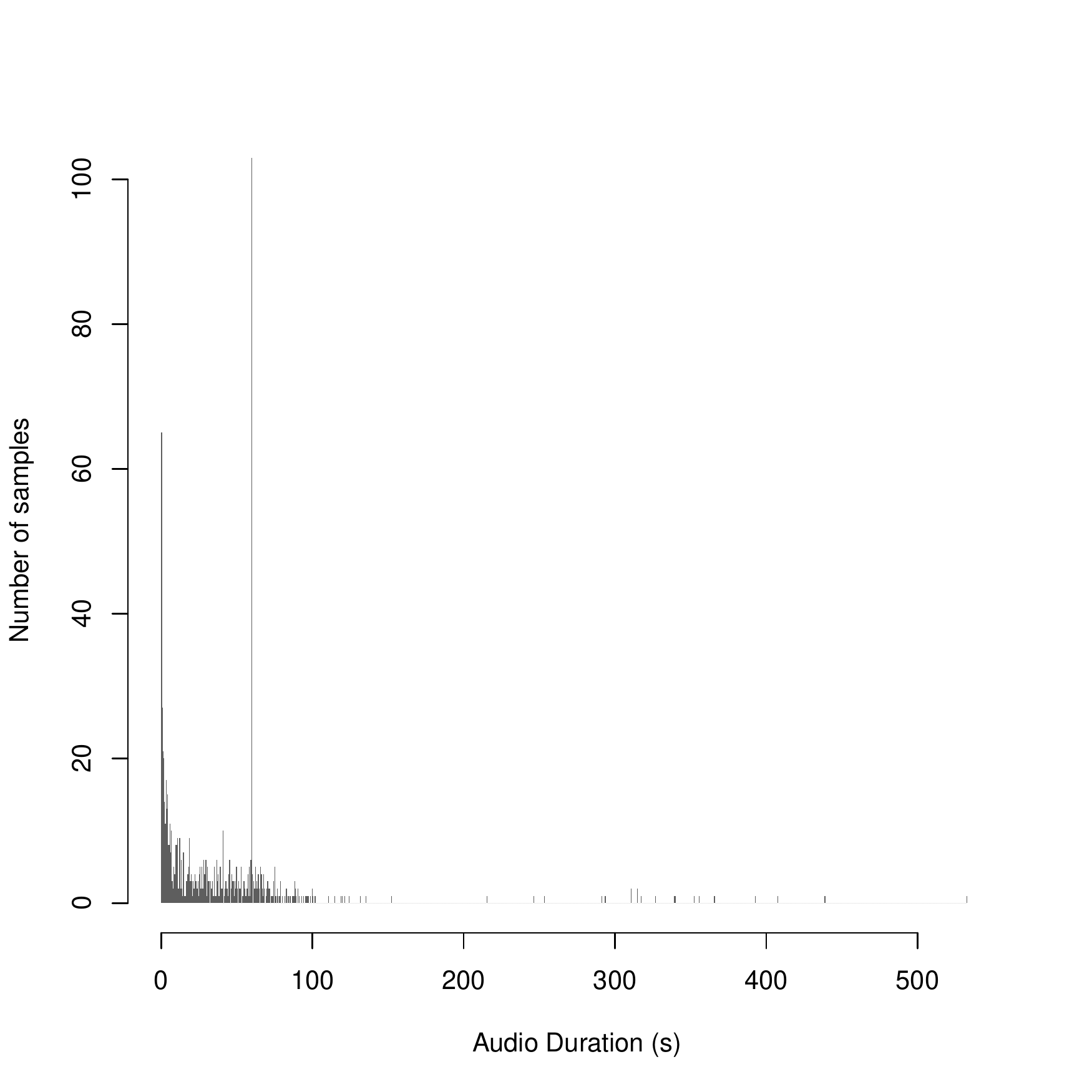}
}%
\quad
\subfigure[]{%
    \centering
    \label{fig:audio:frequency}%
    \includegraphics[width=0.42\columnwidth]{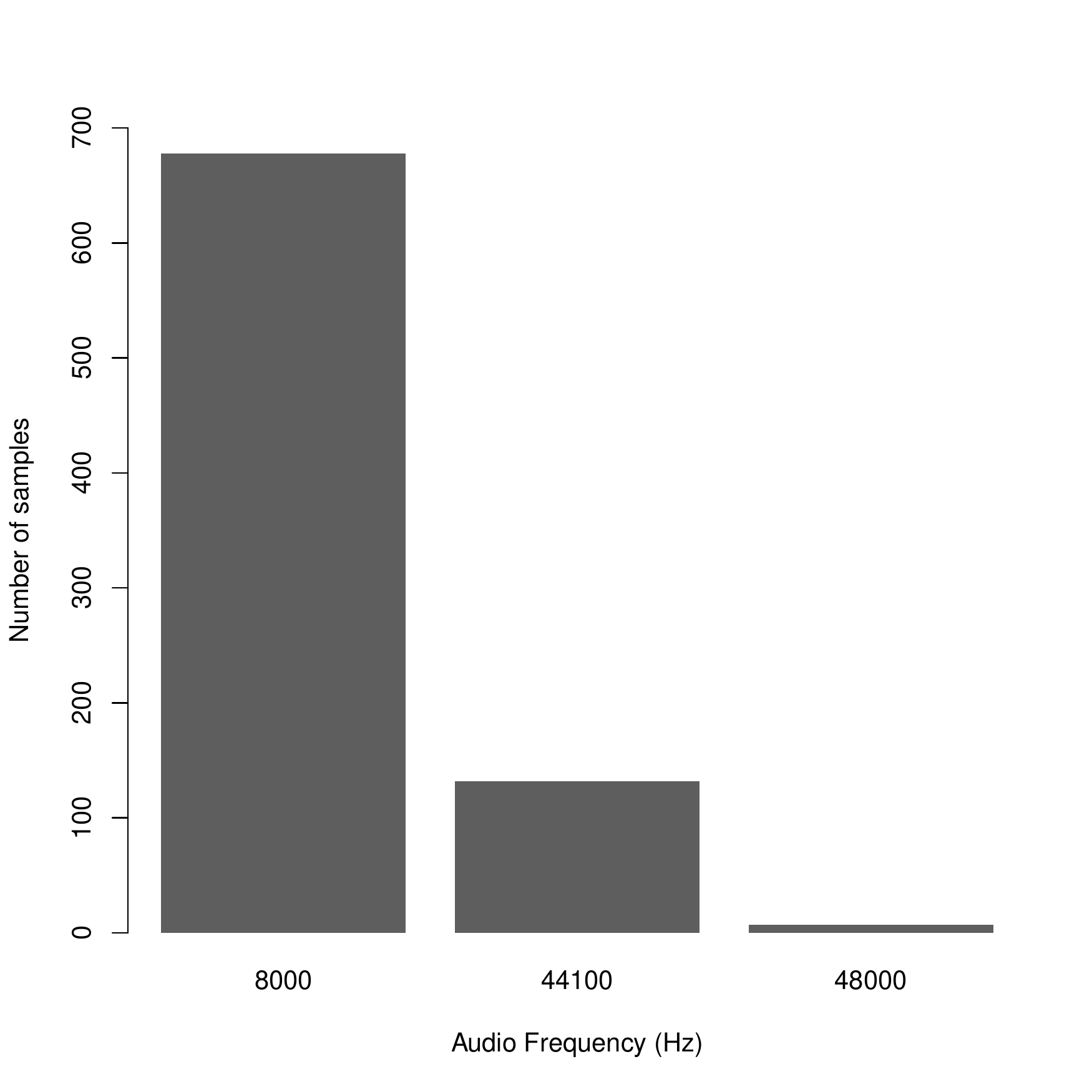}
}%

\caption{Distribution of audio duration and frequency in the original dataset with mosquito wingbeat sounds.}
     \label{fig:audio}
\end{figure}

We downloaded the dataset generated by the Abuzz project \citep{mukundarajan2017using}, which contains 1,285 audio files for 20 species of mosquitoes. Acoustic data on species-specific mosquito wingbeat sounds were recorded using commercially available mobile phones. Table~\ref{tab:audiodata} summarizes the number of recordings and their total duration per specie. We note that the species \textit{Anopheles arabiensis} and \textit{Anopheles gambiae} are each sub-classified in two laboratory-reared strains. 
We performed the dataset analysis using the Python libraries \texttt{librosa} (for audio manipulation), \texttt{numpy} (for scientific computing), and \texttt{matplotlib} (for generation of graphics). The files were recorded using different mobile phones and distinct file formats (WAV, MP4, M4A, and AMR), and most species have recordings generated with two or more mobile phones. As a result, audio duration and sampling frequency vary as can be seen in Figure~\ref{fig:audio}.

To control sources of bias in model training, we pre-processed the files to standardize their format and sampling frequencies. We also converted them to WAV format using the open source software Audacity\footnote{Audacity website: \url{https://www.audacityteam.org/}.}. During pre-processing, we manually removed audio segments without any mosquito sound, as well as segments in which background noise was as loud as the sound produced by the mosquito wingbeat. Manual removal of noisy data was necessary as the sample sizes are limited, such that maintaining any satisfactory audio segment is crucial to improve model fitting to data by ML algorithms. On the other hand, low-quality training data may negatively interfere in the model's generalization power. Therefore, balancing data quality and volume is a relevant issue. 

We also adopted the lowest sampling frequency found in the original files, \ie 8,000 Hz, as a standard in our work. This value is much higher than 700Hz, the maximum frequency of mosquito wingbeats registered in literature, such that the Nyquist–Shannon sampling theorem is being properly respected. 
To change the sampling frequency, we used SoX\footnote{SoX source code: http://sox.sourceforge.net}, a command-line audio processing tool. The duration of sound files for each species after the pre-processing step is shown in Table~\ref{tab:audiodata}. After these pre-processing steps, a certain level of normalization among input files is guaranteed, except for the normalization of the volume, which has not been modified.

\subsection{Features extraction}

Time-frequency representations of sounds are especially useful as learning features in audio classification \citep{huzaifah2017comparison}. 
Following previous works \citep{dieleman2014end,piczak2015environmental,cakir2017convolutional}, features extraction from the mosquitoes wingbeat sounds was based on the spectrograms generated from the sound recordings. Spectrogram is a visual representation of the frequency spectrum over time, commonly generated using the FFT. An advantage of this approach is the conversion of audio data into an image-like data, for which the application of machine learning methods such as deep learning is well consolidated. 

In problems of audio classification using ML, most researchers do not use the raw spectrograms, which have a linear scale, but instead the logarithmic scale or the mel scale, as these reduce the resolution of higher frequencies and better resembles human perception \citep{dieleman2014end}.  
Moreover, an experimental comparative evaluation has shown that mel-scaled  short-time Fourier transform (STFT) transformation outperforms other approaches as time-frequency representations for environmental sound classification \citep{huzaifah2017comparison}. 
 
Therefore, we extracted the mel-scaled spectrogram as a feature using the 
Python \texttt{librosa} library. Different combination of FFT parameters values were tested, varying each parameter within a set of pre-specified values, as follows:  i) number of bands $\in$ \{20,40,60,80\}, ii) number of frames $\in$ \{20,40,60\}, iii) hop length $\in$ \{64,128,256,512\}, iv) window size $in$ \{512, 1028, 2048\}. As our training data contains limited number of samples per specie, we adopted a 50\% overlap among FFT windows to obtain more samples. Values returned by the function are originally in a range from -80 to 0 dB, and were normalized to [0,1] interval using the following equation. 
\begin{equation}
    x_{norm} = \dfrac{x}{80}+1
\end{equation}

\subsection{Convolutional Neural Networks} 
Our approach to audio classification from mosquitoes wingbeats sounds deploys Convolutional Neural Networks (CNNs) trained on mel-scaled spectrograms. CNNs are one of the most popular deep learning networks architectures used in computer vision due to their outstanding ability in detecting patterns from images \citep{rawat2017deep}. Since spectrograms are visual representations of audios, CNNs have been successfully used to extract patterns from them in a similar way to the methodology for image analysis and object recognition \citep{piczak2015environmental,cakir2017convolutional}. 

CNNs are feedforward networks whose architecture, despite several possible variations, has its core based on convolutional and pooling layers grouped into modules in an alternated fashion \citep{rawat2017deep}. Convolutional layers serve as high-level features extractors, applying a convolution kernel over the input and providing a feature map as output. The last component of the convolutional layer is the activation function to increase the non-linearity in the output, for which the rectified linear units (ReLUs) have become popular \citep{nair2010rectified}. Pooling layers are commonly inserted between successive convolutional layers to progressively reduce the spatial size of the data representation (\ie resolution of the feature maps) and help control overfitting \citep{Patterson&Gibson2017}. Modules are often stacked on top of each other after the input layer to form a deep model and extract more abstract feature representations as information propagates through the network. The set of feature-extraction layers are followed by one or more fully connected (FC) layers, which interpret the feature representations and perform the high-level reasoning about the classification of the input \citep{rawat2017deep}. The last fully connected layer outputs the class label and is thus called the classification layer.  

Large complex models as CNNs, comprised by thousands or millions of parameters, are prone to overfitting if trained on relatively small data sets as it is the case in the current work. To prevent large models from overfitting, dropout learning has been widely used as an efficient regularization technique \citep{cai2019effective}. It consists in randomly sampling a set of neurons to be removed from a network layer in each training iteration and performing training in the remaining sub-network, which has the effect of making nodes in the network generally 

\subsection{Model Evaluation}
\label{sec:model_evaluation}

Models were evaluated using stratified 10-fold cross-validation. 
Due to the limited sample size, no hold-out division (\ie dividing original data into a training and a testing set) was applied prior to cross-validation. Traditional evaluation metrics were adopted to assess distinct types of correct prediction, \ie true positive (TP) and true negative (TN), and incorrect predictions, \ie false positive (FP) and false negative (FN), and allow a more detailed analysis of classifiers' predictive power. The following metrics were analyzed:
\begin{itemize}
    \item \textbf{Accuracy}: the proportion of instances in the testing set that were predicted correctly, divided by the total number of instances in the testing set. Although accuracy is a commonly applied metric and easy to interpret, it can become an unreliable measure of model performance for class imbalanced datasets, especially for severe skews in the class distributions.
    \item \textbf{Precision}: the number of positive instances predicted correctly (\ie TP) divided by all positive predictions made by the model. Low precision indicates a high number of false positives. 
    \item \textbf{Recall}:  the number of true positives predicted by the model divided by the total number of positive instances in the testing set. Recall is also called Sensitivity or True Positive Rate, and a low recall indicates a high number of false negatives.
    \item \textbf{F1-measure (F1)}: the harmonic mean between precision and recall, attaching equal weights to both false positives and false negatives.
\end{itemize}

Performance values are presented as means and standard deviations across 10 folds. For multiclass classification problems, performance for each metric is summarized as the macro average across all classes.

\section{Proposed Approach} 
\label{sec:approach}

\begin{figure}[!t]
    \centering
    \includegraphics[scale=0.85]{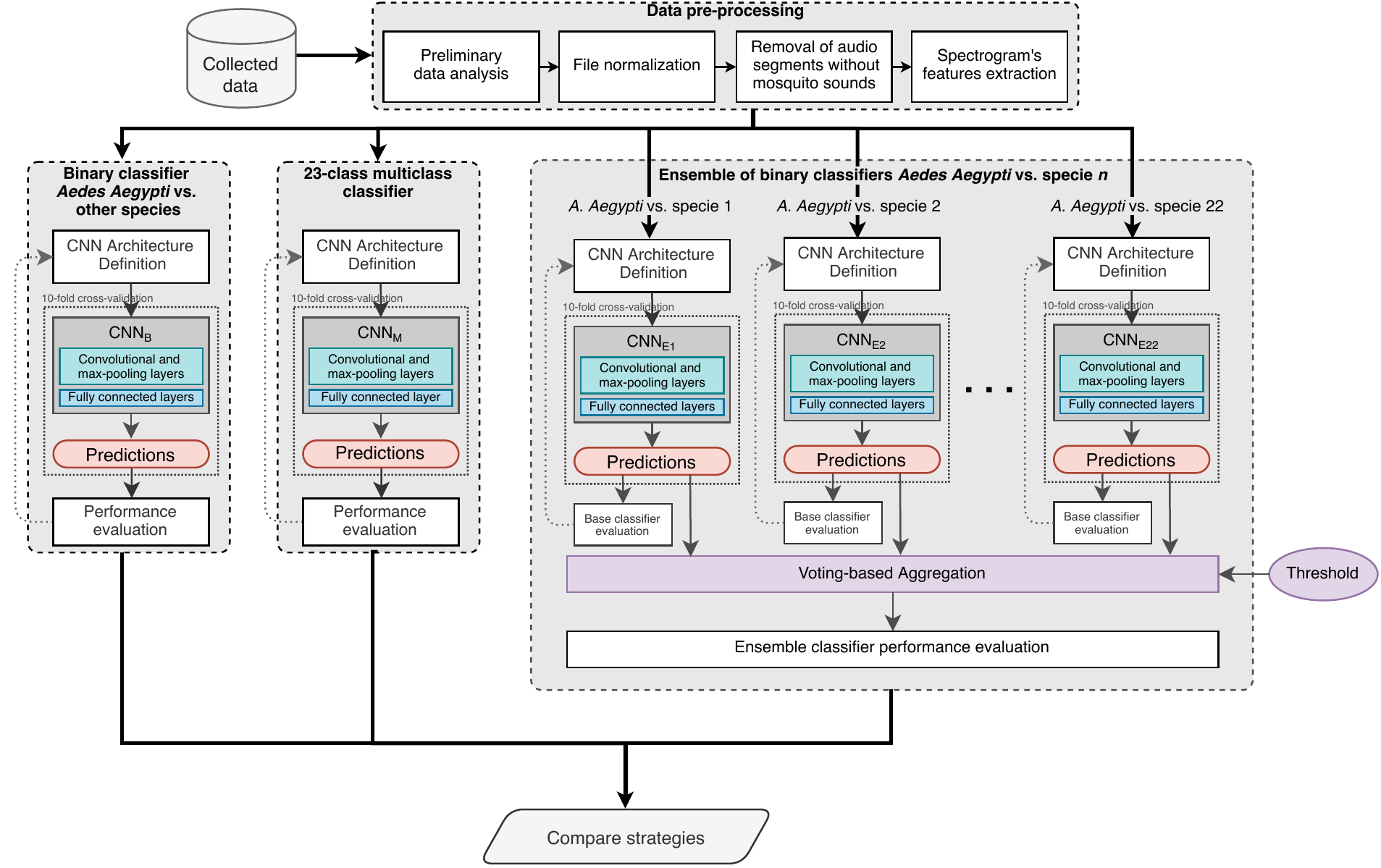}
    \caption{Workflow considered in the analysis of the proposed methodology.}
    \label{fig:workflow}
\end{figure}

In Figure~\ref{fig:workflow} we present an overview of the workflow we considered for developing and assessing the performance of three different strategies for identification of \aedes: using a binary classifier, a multiclass classifier, and an ensemble of binary classifiers. 

\subsection{Binary Classifier} 
\label{sec:approach:binary}

In this strategy, we created a binary model adopting a one-vs-all approach, in which a classifier is trained to distinct between \aedes and all other mosquito species, which were considered as a single class (\ie the negative class). Although extremely unbalanced, this network represents the most straightforward approach to perform the classification. Our goal is to train a model with high sensitivity (\ie recall) for the detection of  \aedes.
                  
The CNN architecture for the binary classifier is composed of three convolutional layers, the first two being followed by a max-polling layer of size 2 x 2 with stride 1. The first convolutional layer applies 32 filters with a 3 x 3 kernel using the ReLU activation function, whereas the second and third convolutional layers apply 64 filters with 3x3 kernels and the ReLu activation. Finally, two FC layers are arranged in sequence: the first one is composed of 256 neurons and the ReLU activation, whereas the second FC, referring to the classification layer, has two output neurons and adopts the Sigmoid activation function.

\subsection{Multiclass Classifier} 
\label{sec:approach:multiclass}

In the second approach, each strain was considered as an individual class and we trained a model to classify a new input sound among the 23 possible classes. Because our focus is on the specific identification of \aedes, we did not investigate further the multiclass strategy. However, it is interesting to note the performance of this model and, in particular, its confusion matrix, as both enable a better understanding of the challenges when trying to classify mosquito species. Also, training this model enables a comparison with the results obtained in the Abuzz project \citep{mukundarajan2017using}, which performs classification of audio recordings using the Maximum Likelihood estimation algorithm. 

The CNN architecture for the multiclass classifier was composed of two modules alternating a convolutional layer and a pooling layer, after which a FC layer was attached. The initial convolutional layer (C1) applies 32 filters with a 20 x 5 kernel using the ReLU activation function, followed by a max-pooling layer of size 2 x 2 with stride 1.  The second convolutional layer (C2) has 32 filters with 8 x 4 kernels and also applies the ReLu activation function. It is followed by a max-pooling scheme with 2x2 filter and stride 1. Finally, the FC layer applies the Softmax activation function to classify the input instance in one of the 23 possible classes.

\subsection{Ensemble of Binary Classifiers} 
\label{sec:approach:ensemble}

Finally, we developed a model that consists of an ensemble composed of 22 binary base classifiers. The base classifiers have in common the \aedes as the positive class in the training data. However, the negative class varies, being a different species of mosquito for each classifier (\ie one of the 22 non-\aedes species). We hypothesize that when we confront the ensemble with \aedes audio, most base classifiers will predict a higher probability for the \aedes class, such that the consensus-based prediction will be correct. On the other hand, when we present another species as input to the ensemble, it is reasonable to think that more divergence will be observed among their outputs and, consequently, the consensus prediction will tend for the negative class.  

To create the ensemble consensus, we decided for a simple combination scheme based on majority voting among the base classifiers. Majority vote counts the votes for each class
over the base classifiers and selects the majority class. We tested and compared different thresholds for the majority voting, here defined as the minimum number of votes required for the positive class so that this is considered as the ensemble prediction. Thresholds were varied starting from the simple majority (\ie minimum of 50\% of votes) to a strictly majority decision based on a 95\% threshold.

Base classifiers were trained using the same CNN architecture adopted for the binary classification model, as described in Section~\ref{sec:approach:binary}. The cross-validation process was adapted to ensure that for a given fold (i) the same instances of \aedes are used to train all base classifiers and (ii) the same testing set is used to evaluate all the base classifiers (\ie containing test instances from all the species), avoiding overlap between training and testing data for each single base classifier. Therefore, in each fold, we created a stratified testing set with a portion of instances from the 23 species, and 22 training sets composed by the remaining instances from \aedes and each non-\aedes species.

\section{Experiments and Results}
\label{sec:experiments}


To test the proposed strategies, we used Keras\footnote{Keras website: \url{https://keras.io/}.} 
to train their respective CNN models, whose architecture were described in Section~\ref{sec:approach}. The number of iterations (\ie epochs) and batch size were chosen based on empirical evaluation, by analyzing the accuracy and error of the training set and validation throughout the interactions. Ten epochs and a batch size of 32 were applied in our experiments. 

We adopted the categorical cross-entropy loss function for model compilation, defined as:
\begin{equation}
\mathrm{Loss} = -\sum_{i=1}^{\mathrm{C}} y_i \cdot \mathrm{log}\; {\hat{y}}_i
\end{equation}
where C is the number of classes (or the number of scalar values in the model output), \(\hat{y}_i\) is the \(i\)-th scalar value in the model output, and \(y_i\) is the corresponding target value.

CNNs hyperparameters were tuned using the Root Mean Square Propagation (RMSProp) optimizer. 
Stratified 10-fold cross-validation was implemented for model validation with the Python \texttt{scikit-learn} library \citep{pedregosa2011scikit}. 

In this section, we first present the results from the experimental analysis considering different configurations of FFT parameters over models' performance (Subsection~\ref{subsec:analysis-fft}). Then, we provide a detailed analysis of performance indicators, regarding the accuracy, precision, recall, and F1-measure, for all the binary (Subsection~\ref{subsec:performance-binary}), multiclass (Subsection~\ref{subsec:performance-multiclass}), and an ensemble of binary classifiers (Subsection~\ref{subsec:performance-ensemble}) classification strategies .

\begin{table}[!t]
    \centering
    \resizebox{0.8\columnwidth}{!}{  
    \begin{tabular}{c|c|c|c|c} \hline \hline
         Model config. & No. of bands & No. of frames & Hop length & Window size\\ \hline \hline
       1 & \cellcolor{lightgray} 20 & 40 & 128 & 1024 \\
       2 & \cellcolor{lightgray} 40 & 40 & 128 & 1024 \\
       3 & \cellcolor{lightgray} 60 & 40 & 128 & 1024 \\
       4 & \cellcolor{lightgray} 80 & 40 & 128 & 1024 \\ \hline
       5 & \cellcolor{vlgray} 60 & \cellcolor{lightgray} 20 & 128 & 1024 \\
       6 & \cellcolor{vlgray} 60 & \cellcolor{lightgray} 60 & 128 & 1024 \\ \hline
       7 & \cellcolor{vlgray} 60 & \cellcolor{vlgray} 40 & \cellcolor{lightgray} 64 & 1024 \\
       8 & \cellcolor{vlgray} 60 & \cellcolor{vlgray} 40 & \cellcolor{lightgray} 256 & 1024 \\
       9 & \cellcolor{vlgray} 60 & \cellcolor{vlgray} 40 & \cellcolor{lightgray} 512 & 1024 \\ \hline
       10 & \cellcolor{vlgray} 60 & \cellcolor{vlgray} 40 & \cellcolor{vlgray} 256 & \cellcolor{lightgray} 512 \\
       11 & \cellcolor{vlgray} 60 & \cellcolor{vlgray} 40 & \cellcolor{vlgray} 256 & \cellcolor{lightgray} 2048
       \\ \hline \hline
    \end{tabular}}
    \caption{Combination of FFT parameters analyzed, each combination being a model configuration. The darker gray cells highlight how we varied parameter setting across configurations, and lighter gray cells indicated the most favorable parameter value found in our experiments and used in subsequent configurations.} 
    \label{tab:eval:list-configurations}
\end{table}


\subsection{Analysis of Fourier Transform Parameters}
\label{subsec:analysis-fft}

In this analysis, we assessed the influence of FFT parameters on classification performance to identify which model configurations are more suitable for use with the proposed approaches. Due to the large number of classification tasks, as well as the many details involved in the architecture and learning hyperparameters of CNNs, it was unfeasible to carry out this investigation for every CNN model trained. For this reason, we performed the analysis of different configurations of FFT parameters using a binary classifier whose goal is distinguishing \aedes from \emph{Anopheles freeborni}; the parameter setting used in the model configuration with best performance were then transposed to the other classifiers. We chose \emph{Anopheles freeborni} as the negative class because of its similarity in terms of audio duration to \aedes, which helped minimize the effect of class imbalance. We set up the CNN architecture as described in Section~\ref{sec:approach:binary}, and varied the model configuration in terms of FFT parameters as shown in Table~\ref{tab:eval:list-configurations}. In the following plots and discussion, we use \emph{config. \#n} (or \#n for short) as a reference to the model evaluated using the n-th set of parameters, as listed in the column \emph{model config.}

\begin{figure}[h]
    \centering
\subfigure[]{%
    \centering
    \label{fig:eval:fftconfigurations:accuracy}%
    \includegraphics[width=0.48\columnwidth]{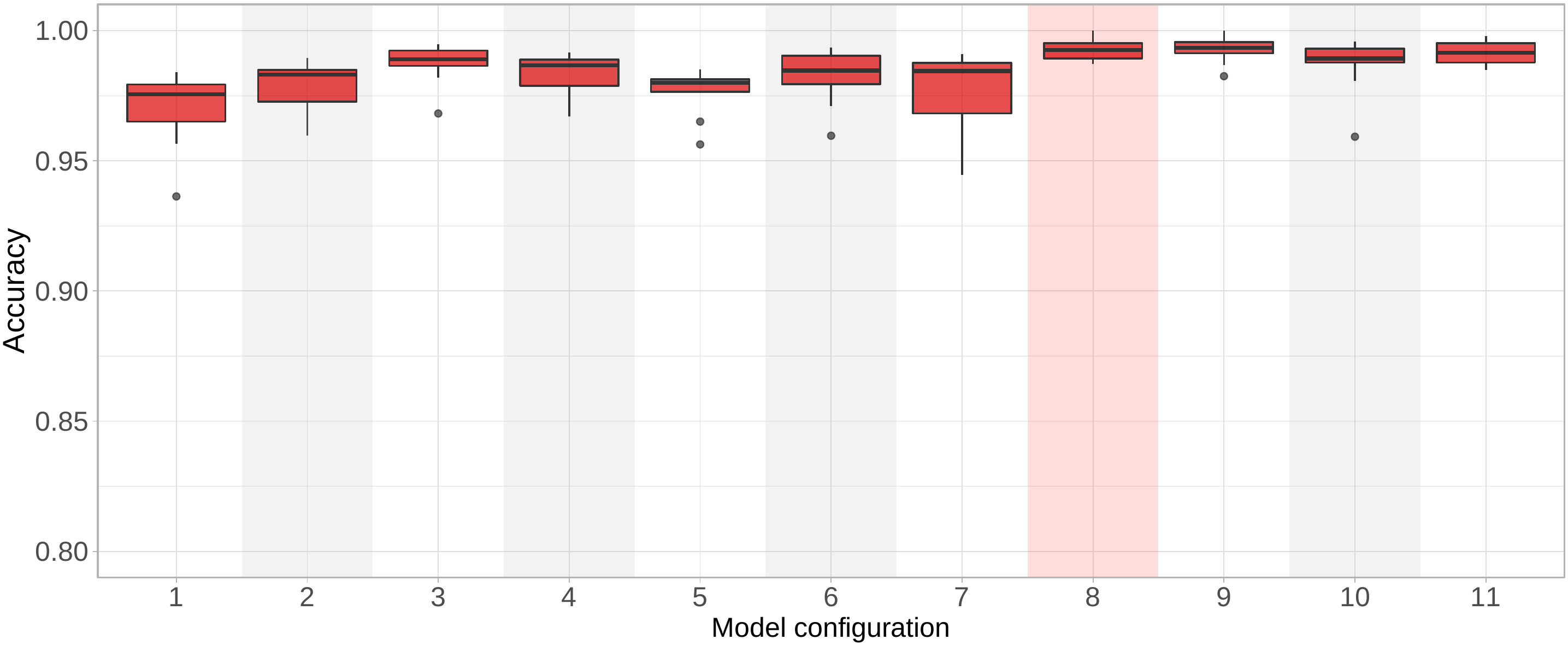}
}%
\quad
\subfigure[]{%
    \centering
    \label{fig:eval:fftconfigurations:precision}%
    \includegraphics[width=0.48\columnwidth]{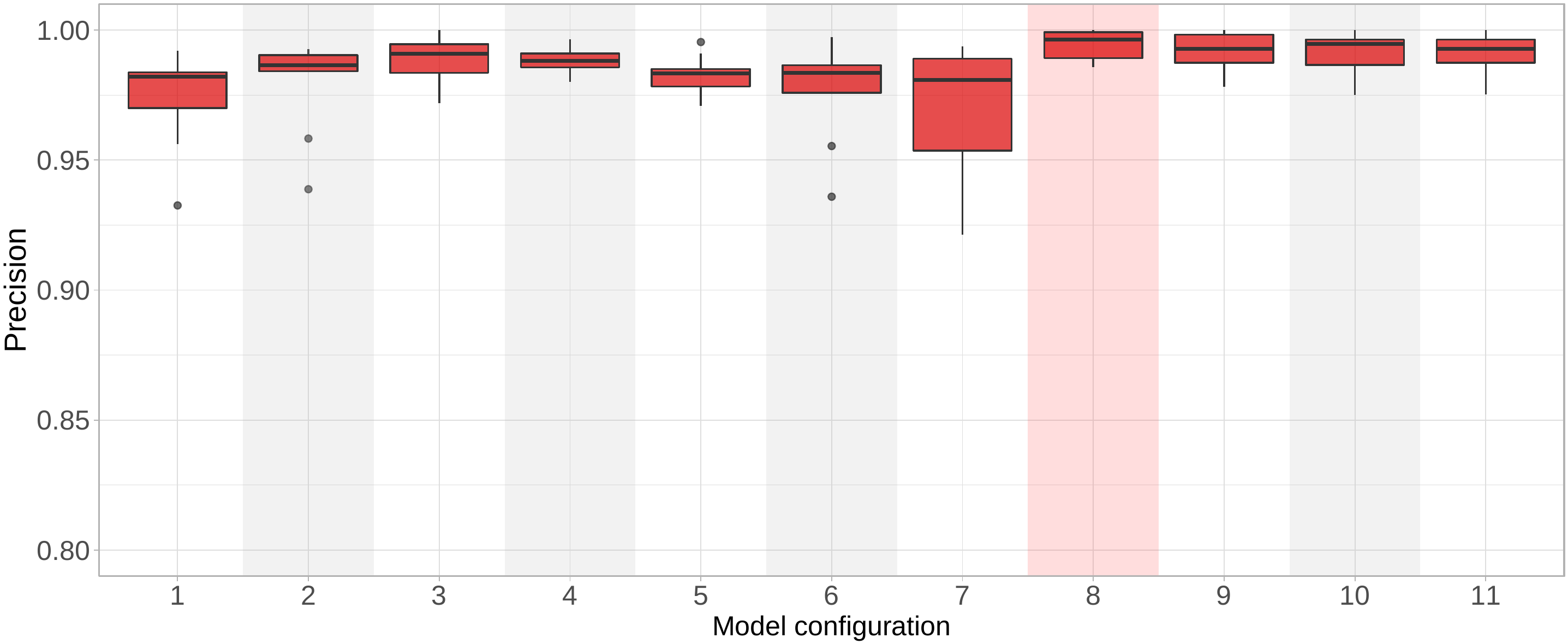}
}%
\quad
\subfigure[]{%
    \centering
    \label{fig:eval:fftconfigurations:recall}%
    \includegraphics[width=0.48\columnwidth]{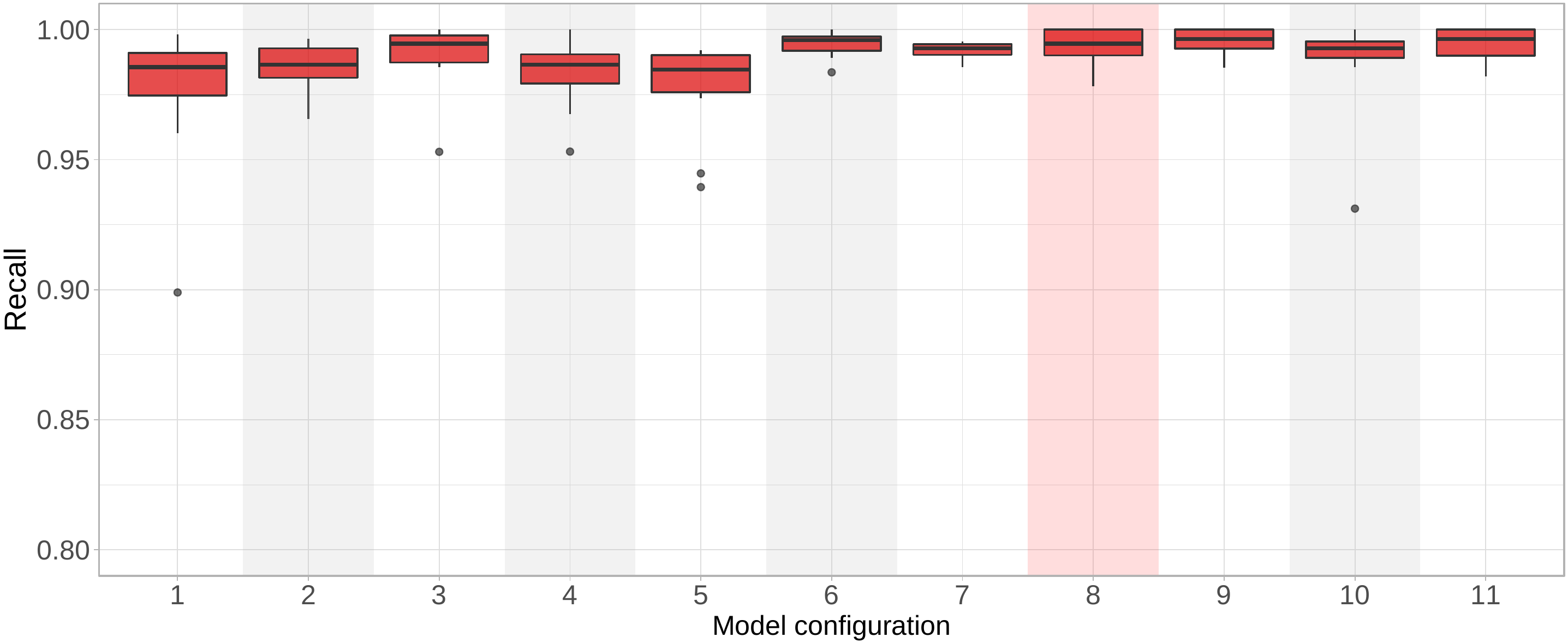}
}%
\quad
\subfigure[]{%
    \centering
    \label{fig:eval:fftconfigurations:f1measure}%
    \includegraphics[width=0.48\columnwidth]{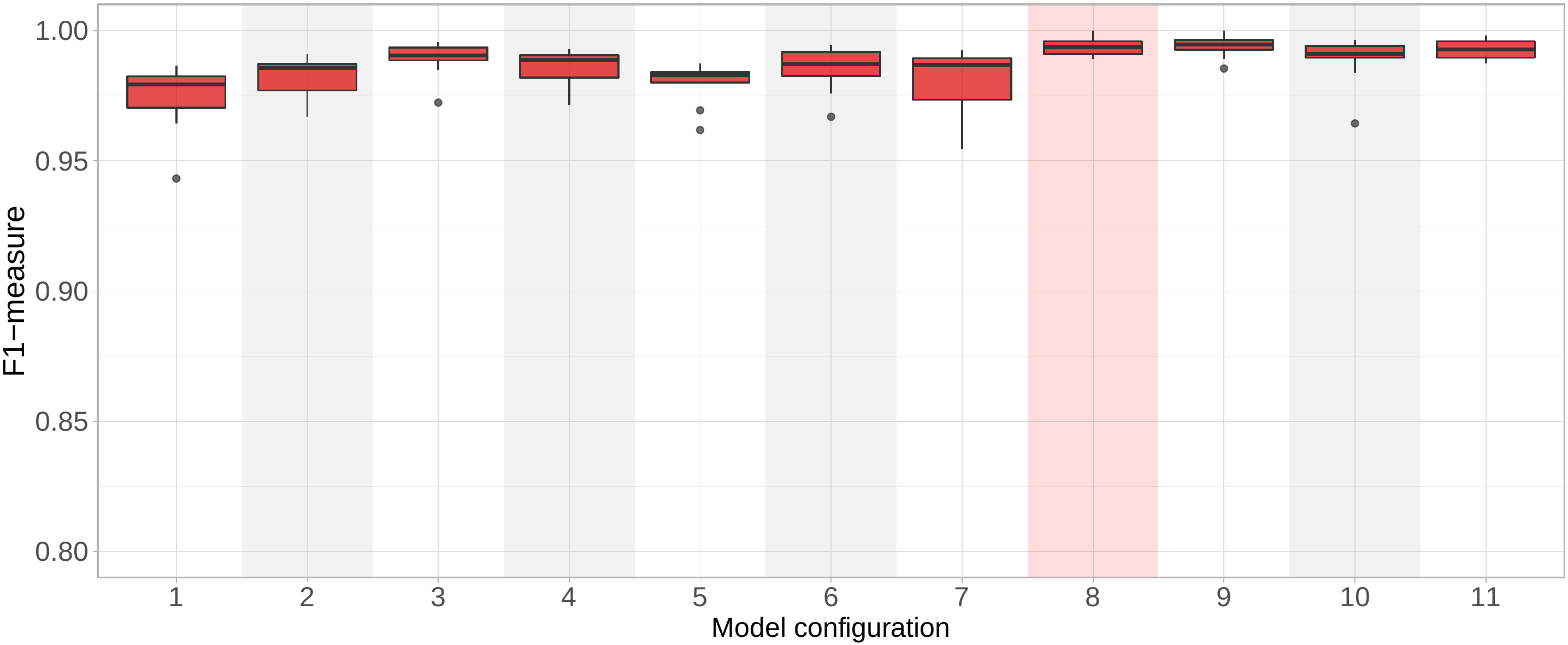}
}%
\caption{Performance distribution of accuracy, precision, recall, and F1-measure for the model configurations tested (Table~\ref{tab:eval:list-configurations}). In summary, config. \#8 presented the best overall performance and was selected for further experiments.}
    \label{fig:eval:fftconfigurations}
\end{figure}

Figure~\ref{fig:eval:fftconfigurations} presents the performance distributions for this experimental analysis. We obtained the samples from a 10-fold cross-validation process. 
Overall, the trained models presented high performance, with most metrics achieving a predictive score higher than 0.9, except for some outliers. 
We note the impact of changes in the FFT parameters over the classifier performance. In config. \#1 to \#4, we varied the number of bands: 20, 40, 60, and 80, respectively. The number of frames, hop length, and window size were initially set to 40, 128, and 1024, respectively, and varied in subsequent configurations (as discussed in the remainder of this subsection). Observe in Figure~\ref{fig:eval:fftconfigurations} that the measured model performance regarding accuracy (a), precision (b), recall (c), and F1-measure (d) increases with the number of bands, reaching the most stable performance with 60 bands (config. \#3). Then, with 80 bands, there is a slight degradation in the model performance, suggesting that 60 bands are most suitable for the binary classifier analyzed.

For the subsequent model configurations, we fixed the number of bands in 60. We then varied the number of frames for config. \#5 and \#6: 20 and 60 frames, respectively; note that \#3 already contemplates 40 frames. In summary, config. \#3 remained the one with better performance; \#5 presented a significantly lower recall (with more outliers), whereas \#6 had worse accuracy and precision. We thus considered 60 bands and 40 frames for the following model configurations.


In config. \#7 to \#9, we varied the hop length: 64, 256, and 512; again, note that config. \#3 contemplates hop length 128. In this case, we obtained the best overall performance for recall with hop length 128 and 512, and the most balanced performance (\ie, with the best trade-off between FP and FN) with config. \#8 (256 frames). Finally, we assessed the effect of varying window sizes in config. \#10 to \#11: 512 and 2048; config. \#8 contemplates a window size of 1024. In this case, observe that there is no significant impact over average performance, although there are more outliers for a window size of 512.

\begin{figure}[h]
 \centering
\subfigure[]{%
    \centering
    \label{fig:eval:fftconfigurationsMean:avg}%
    \includegraphics[width=0.48\columnwidth]{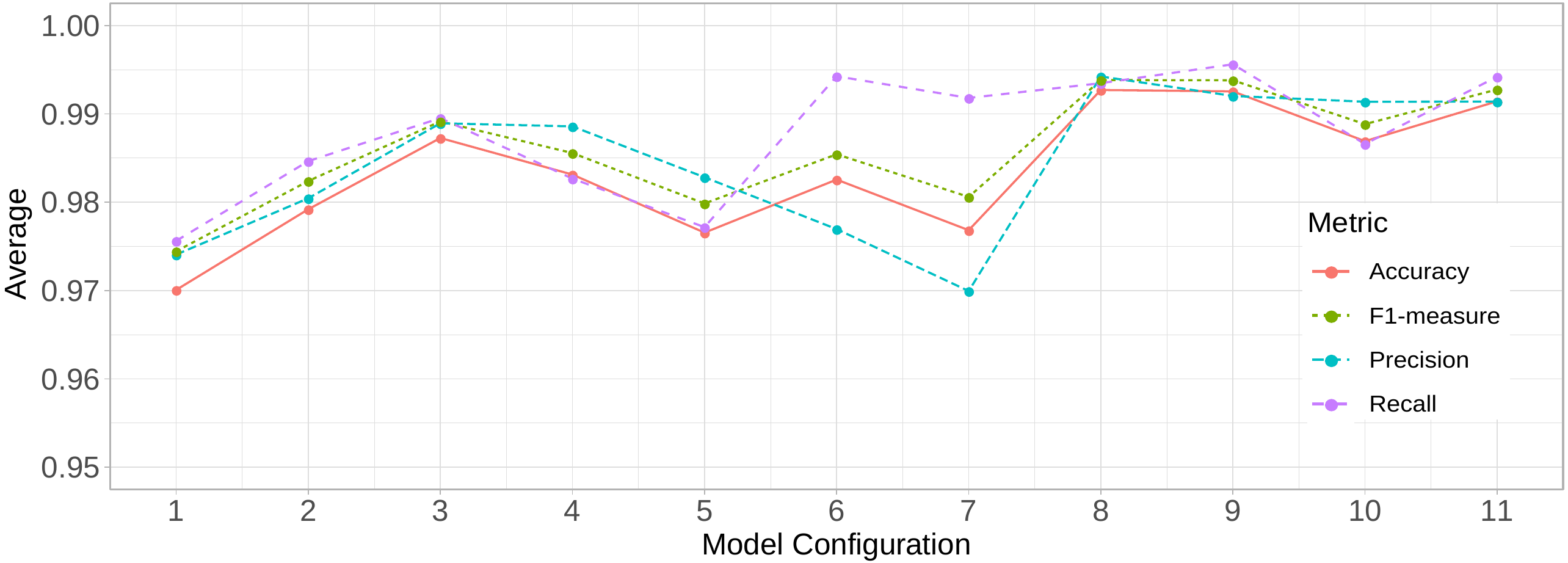}
}%
\quad
\subfigure[]{%
    \centering
    \label{fig:eval:fftconfigurationsMean:stddev}%
    \includegraphics[width=0.48\columnwidth]{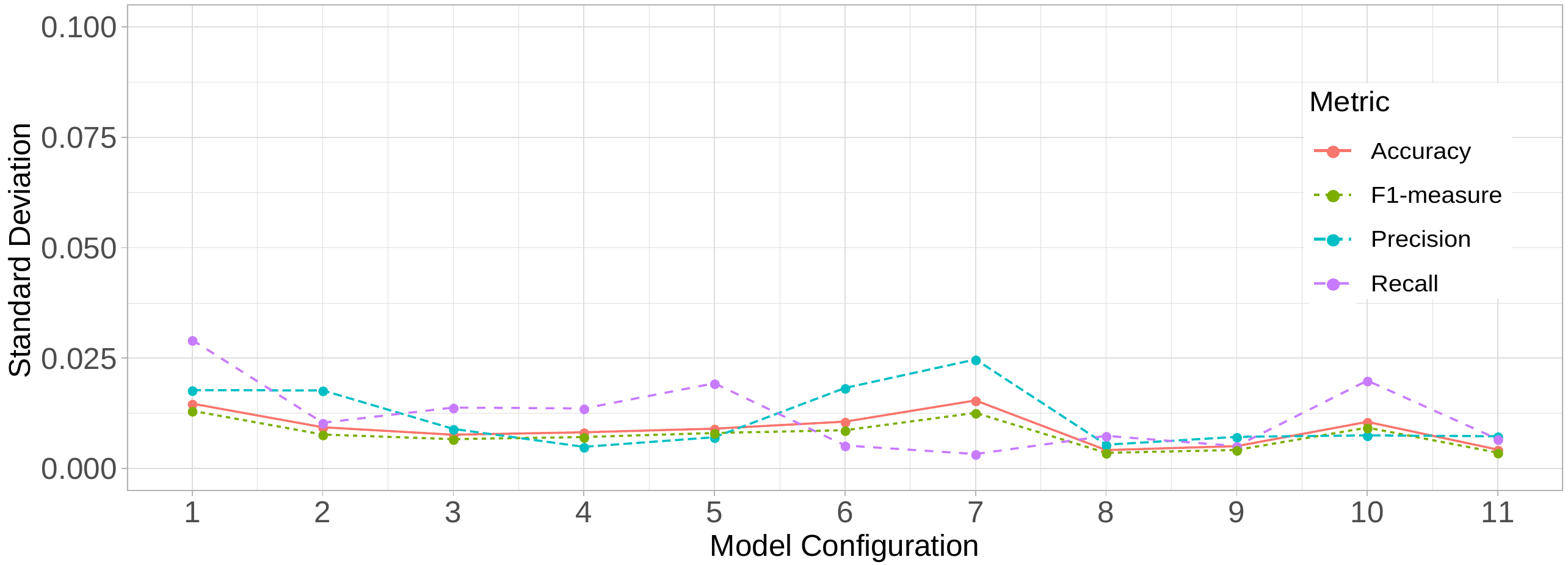}
}%
 \caption{Average and standard deviation for performance metrics analyzed for each model configuration.}
  \label{fig:eval:fftconfigurationsMean}
\end{figure}

Finally, Figure~\ref{fig:eval:fftconfigurationsMean} shows the average performance and standard deviation for each of the metrics considered in our analysis. This figure highlights the most stable performance and the smallest standard deviation obtained for config. \#8. For these reasons, we used the parameter setting in config. \#8 in the experiments discussed next.

\subsection{Performance analysis of the binary classifier}
\label{subsec:performance-binary}

Our first deep learning approach for \aedes detection was based on a binary classifier, which learns to distinguish \aedes from all other species considered as a single negative class (\ie non-\aedes). 
We trained the binary classifier using model configuration \#8 for FFT parameters and a CNN architecture as described in Section~\ref{sec:approach:binary}. 
The training data for building this model presents a severe class imbalance,  with only 11.14\% of instances representing \aedes sounds \ie belonging to the positive class. 
Therefore, developing a model without a predictive preference for the majority class, in this case the negative class, is a natural challenge in this scenario. 

\begin{figure}[h]
    \centering
    \includegraphics[width=0.6\textwidth]{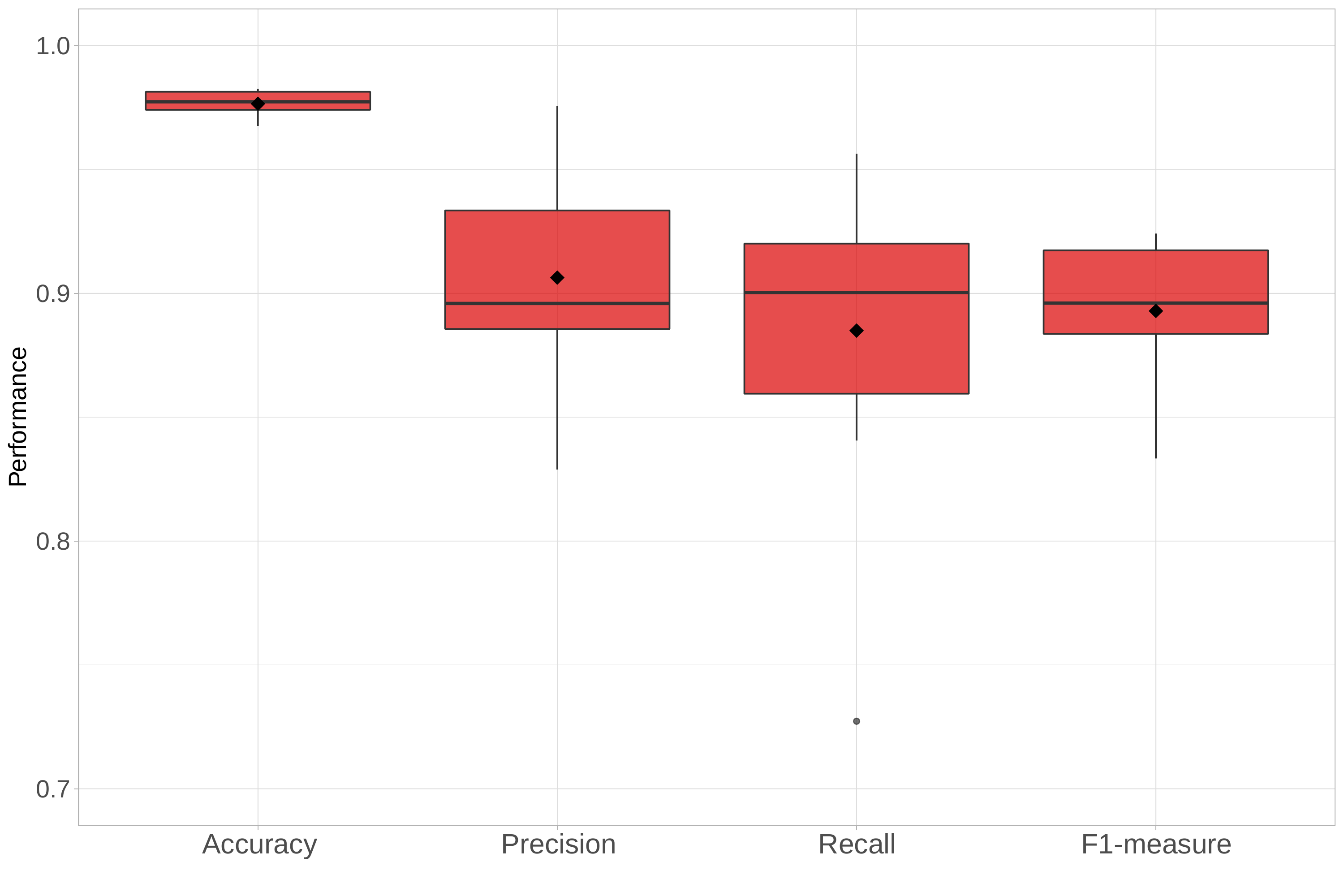}
    \caption{Performance distribution for the binary model using FFT parameters configuration \#8. The black diamonds indicate the mean for each performance metric.}
    \label{fig:eval:binaryModelDist}
\end{figure}

The average accuracy for this classifier computed across the 10 folds of the cross-validation process was 97.65\% ($\pm$0.55) (Figure~\ref{fig:eval:binaryModelDist}). However, we note that as the training data is highly imbalanced in terms of classes distribution, accuracy metric may be misleading in this case. In particular, we highlight the lower average (and higher standard deviation) for F1-measure and its components, precision and recall, as compared to accuracy. Whereas this classifier presents a relatively good precision (90.63\% $\pm$ 4.33), it is not able to recover all true positive labels in the dataset, thus presenting a lower recall (88.49\% $\pm$ 6.68). Therefore, the outstanding mean for accuracy is influenced by the model's ability to catch more TN instances.

The average confusion matrix for the binary classifier is shown in Table~\ref{tab:eval:conf-matrix-binary}. In each cell, we summarize the mean (and standard deviation) for the number of instances classified by the model in each of the four possible groups (\ie TP, TN, FP, and FN). The better predictive performance for the negative class becomes clearer when analyzing the hit rate by class. The average true positive rate (equivalent to recall) was 88.49\% ($\pm$6.68), whereas the average true negative rate was 98.81\% ($\pm$0.61). Thus, the classification model is more accurate and stable for non-\aedes instances. This emphasizes the bias caused by class imbalance towards the negative class, although the capacity for detection of the \aedes was considered to be satisfactory, with medians above 90\%.

\begin{table}[]
\centering
\begin{tabular}{c|c|c|c}
\hline \hline
                            & \multicolumn{3}{c}{Predicted label}      \\ \hline
\multirow{3}{*}{True label} &          & \aedes     & non-\aedes       \\ \cline{2-4} 
                            & \aedes & 243.9 ($\pm$18.44) & 31.7 ($\pm$13.62)   \\ \cline{2-4} 
                            & non-\aedes & 26.3 ($\pm$18.37)   & 2170.9 ($\pm$13.50) \\  \hline \hline
\end{tabular}
    \caption{Average confusion matrix for the binary classifier. Values in each cell refer to the average number of instances (and standard deviation) predicted as (first row) TP, FN, and (second row) FP, TN, across the 10-fold cross validation.}
\label{tab:eval:conf-matrix-binary}
\end{table}

\subsection{Performance analysis of the multiclass classifier}
\label{subsec:performance-multiclass}

In the second proposed approach, we experimented training a CNN model to perform multiclass classification among all mosquito species. This was motivated by the severe class imbalance faced by the binary classifier and by the fact that considering non-\aedes species individually when training the model may improve our understanding of confounding factors and the reasons for misclassified samples. Moreover, previous studies aimed to classify multiple species of mosquitoes (\cite{mukundarajan2017using,fanioudakis2018mosquito}), thus allowing a comparison of our results with data reported in the literature. The CNN architecture was defined as described in Section~\ref{sec:approach:multiclass}, with the FC layer configured with 23 output neurons, one for each possible mosquito species.

In the multiclass approach, \aedes becomes the most frequent among all classes (\ie 11.14\% of instances), followed by \emph{Culiseta incidens} (7.86\%) and \emph{Anopheles freeborni} (7.69\%). The smallest class is \emph{Aedes mediovittatus}, corresponding to only 0.32\% of instances in the dataset. Thus, class imbalance remains a characteristic of the training data, although less detrimental to our study, given that \aedes no longer represents the minority class. 

\begin{figure}[h]
 \centering
\subfigure[]{%
    \centering
    \label{fig:eval:multiClass:avg}%
    \includegraphics[width=0.7\columnwidth]{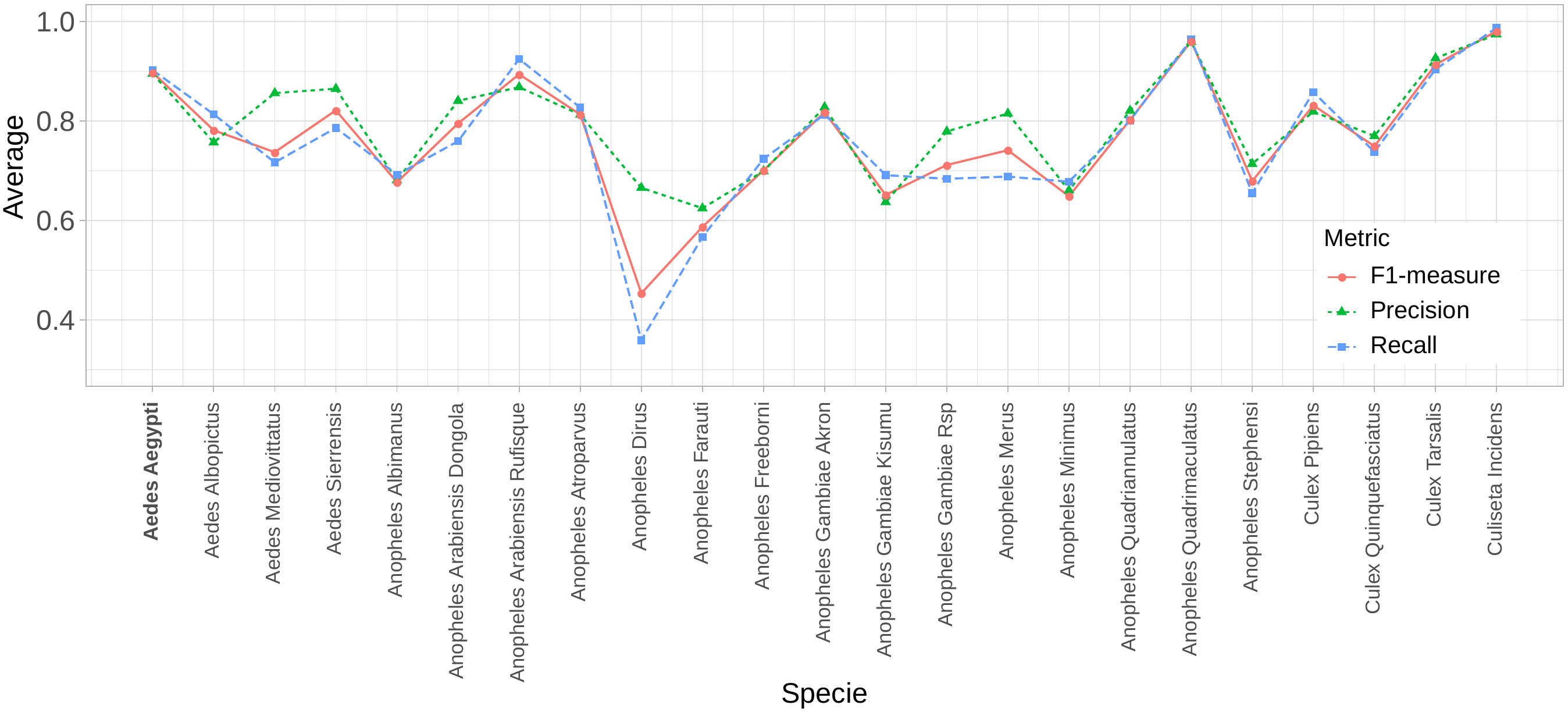}
}%
\quad
\subfigure[]{%
    \centering
    \label{fig:eval:multiClass:stddev}%
    \includegraphics[width=0.7\columnwidth]{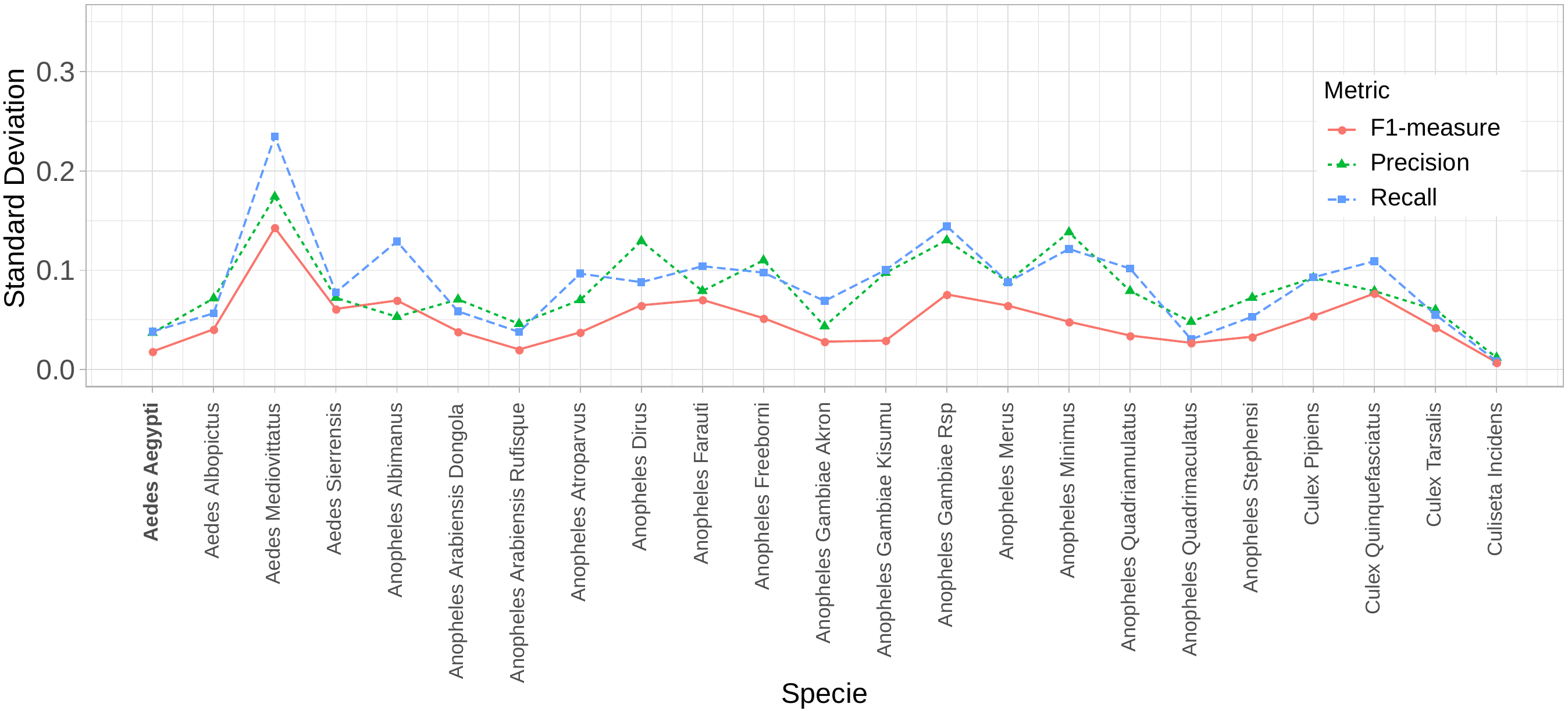}
}%
    \caption{Average and standard deviation performance for multiclass classification model based on configuration \#8.}
     \label{fig:eval:multiClass}
\end{figure}

\begin{figure}[h]
 \centering
\subfigure[]{%
    \centering
    \label{fig:eval:multiclassc8:precision}%
    \includegraphics[width=0.48\columnwidth]{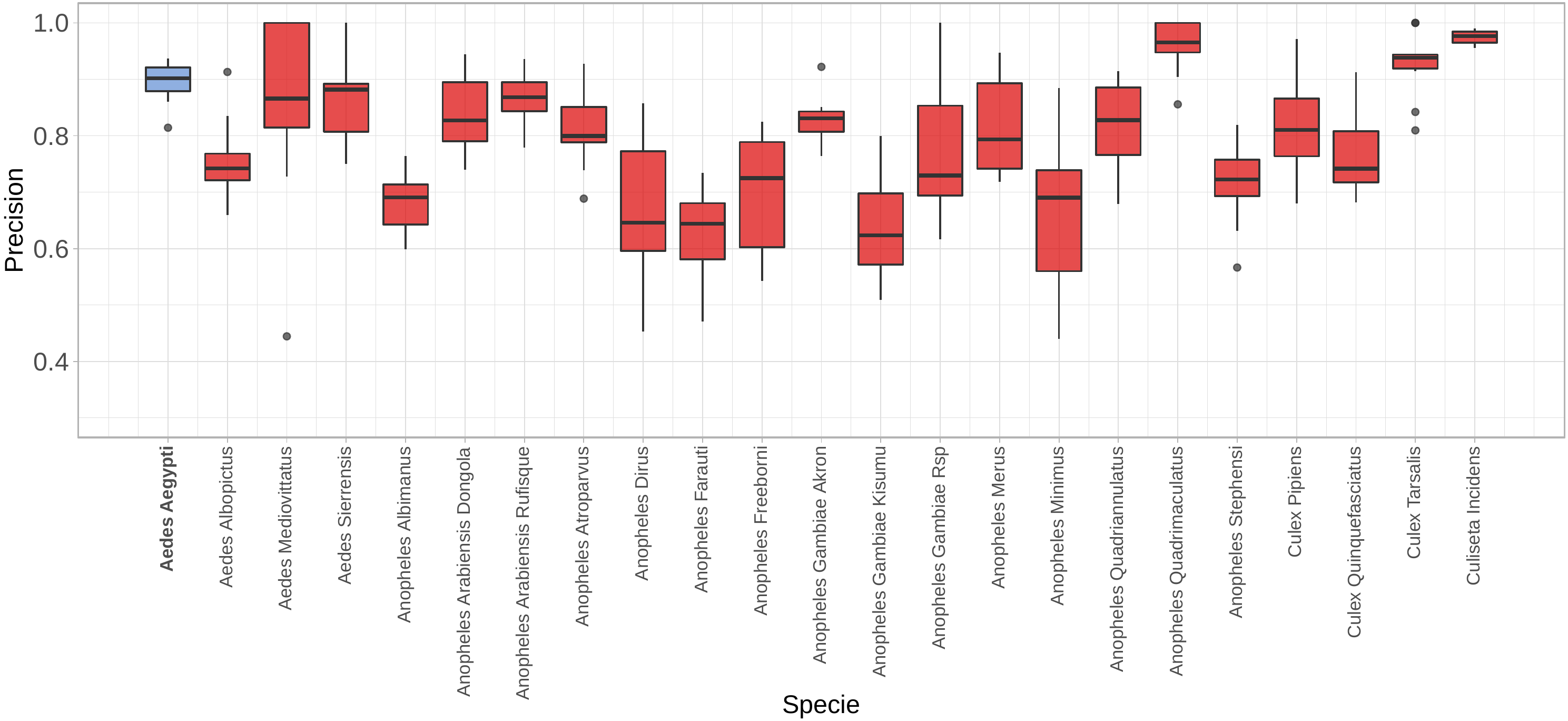}
}%
\quad
\subfigure[]{%
    \centering
    \label{fig:eval:multiclassc8:recall}%
    \includegraphics[width=0.48\columnwidth]{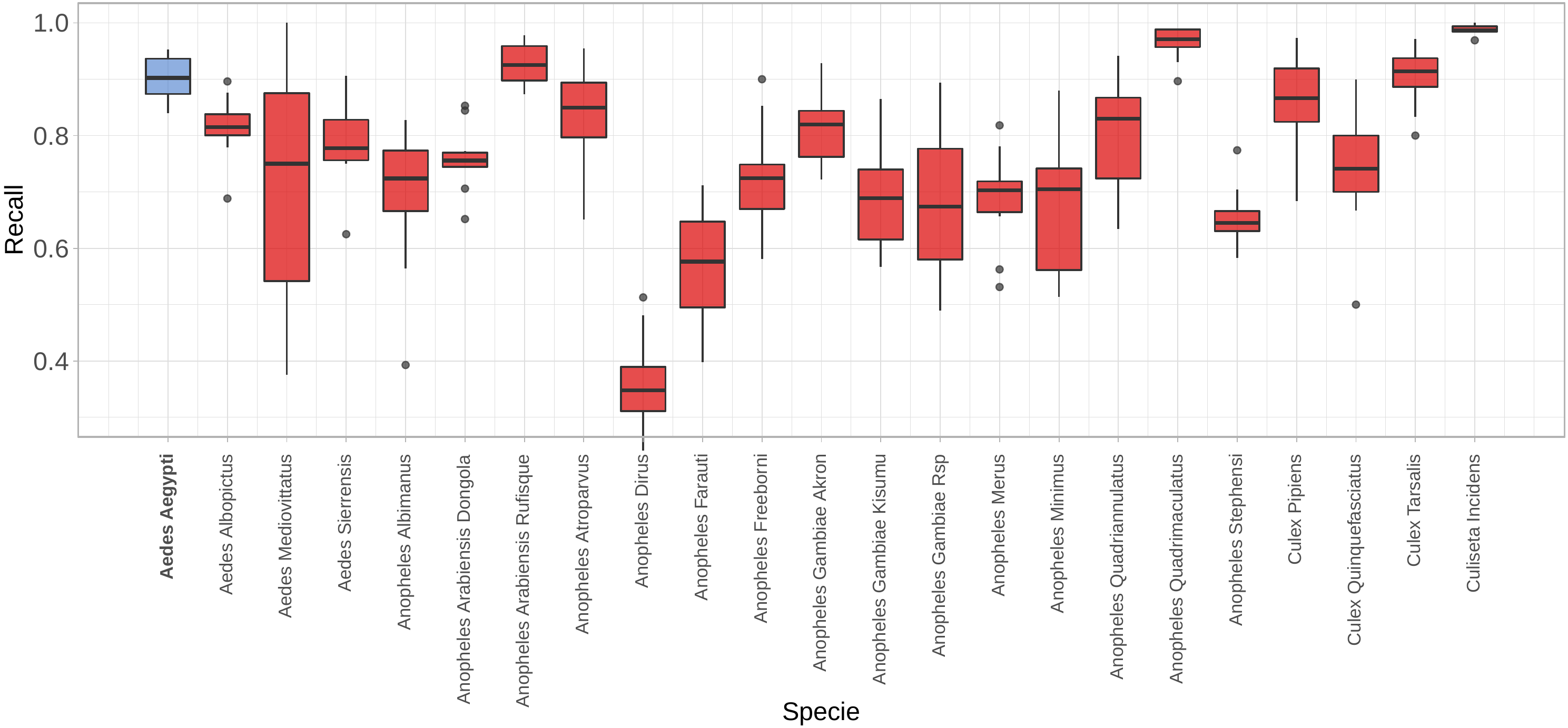}
}%
    \caption{Performance distribution for precision and recall obtained for each species based on the CNN multiclass model. \aedes, our class of interest, is highlighted in the plot.}
    \label{fig:eval:multiclassc8}
\end{figure}

The overall performance achieved by the multiclass CNN model was 78.12\% ($\pm$2.09) of accuracy, 76.71\% ($\pm$13.24) of F1-measure, 79.41\% ($\pm$13.00) of precision, and 76.22\% ($\pm$16.49) of recall. Figure~\ref{fig:eval:multiClass} presents the average (a) and the standard deviation (b) of the CNN multiclass model for each species, considering precision, recall, and F1-measure. The more robust and stable performance for \aedes and \emph{Culiseta incidens}, which are the two most represented species in the training data, is clear: not only the trade-off between precision and recall is good, but also the standard deviation is lower than 5\% for all three metrics. A good performance was also observed for \emph{Anopheles quadrimaculatus}, which corresponds to 3.51\% of training instances. It should be noted, however, that this species has a small number of files in the original dataset collected from a single colony, which may decrease audio variability since only intra-colony variation is reflected.
Of note, model's sensitivity for \aedes was 90.23\% ($\pm$ 3.83), presenting a slight improvement as compared to the binary classifier in terms of its power to detect this particular specie.  

The performance distribution for each species in terms of precision (a) and recall (b) 
is presented in Figure~\ref{fig:eval:multiclassc8}. The variation in performance across species emphasizes that this is not a suitable classifier for the overall task of identifying mosquitoes species by their wingbeat sound, since the predictive power is low in some cases. For instance, the median recall for \emph{Anopheles Dirus} and \emph {Anophelhes Farauti} is lower than 60\%, indicating a high frequency of false negatives in the analysis. In fact, only 8 out of the 22 species have a high detection rate, with an interquartile range for recall above 80\%.  In contrast, the false positive rate is relatively high for several species. Nonetheless, taking as a basis the goal of detecting specifically \aedes, this approach seems to be promising.

\begin{figure}[h]
    \centering
    \includegraphics[width=\textwidth]{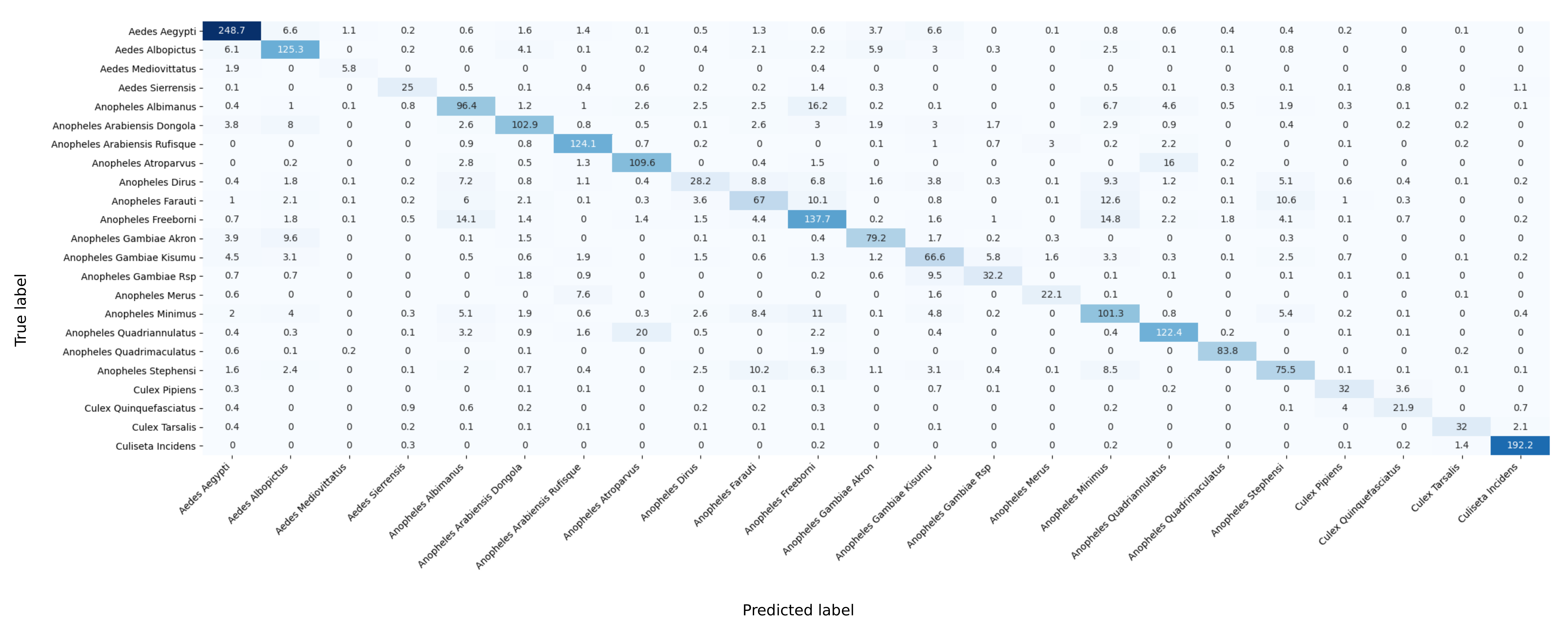}
    \caption{Confusion matrix for the CNN multiclass model, with true labels disposed in the rows and predicted labels provided in the columns. Values within each cell represent the average number of classified samples in the cross-validation.}
     \label{fig:eval:multiClassConf}
\end{figure}

The confusion matrix resulting from the multiclass CNN model is provided in Figure~\ref{fig:eval:multiClassConf}, with true labels disposed in the lines and predicted labels given in the columns. Each cell represents the average number of classified samples in cross-validation. The concentration of higher values within the matrix diagonal is noticeable and more clearly reflects the overall predictive power of the model discussed above. The interesting application of the confusion matrix is to understand the main confounding factors for this model. We note, for instance, that \aedes is often misclassified as \emph{Aedes albopictus} and \emph{Anopheles gambiae} (\emph{Kisumu} and \emph{Akron} strains). Similarly, many instances of these species were incorrectly classified as \aedes by the model. According to \cite{mukundarajan2017using}, \aedes presents a significant overlap in wingbeat frequency distributions with these species, justifying the model's difficulty in carrying out proper classification only based on sound analysis. The distinction between \aedes and \emph{Anopheles gambiae} could be partially enhanced by incorporating other types of evidence, such as recording time, since these species show different circadian rhythms owing to differences in temporal niche, \ie \emph{Anopheles gambiae} is strictly night active and \aedes is primarily day active \citep{rund2013extensive}. We also note that the comparison between \emph{Aedes albopictus} and \aedes is particularly challenging since, besides the overlapping frequency distributions, these species show geographical co-occurrence and similar morphology. Thus, even with geographical location and other metadata, the complete differentiation between these species may be hard to accomplish.

Wingbeat frequency distribution also may explain, at least partially, the robust performance for \emph{Culiseta incidens}. This species not only has a large number of instances in the training set, but it also presents the smallest range of values for its frequency distribution and with a low overlap with the distributions of other species. The overlap occurs mainly concerning \emph{Culex tarsalis}, for which the model incorrectly classifies a small portion of instances of \emph{Culiseta incidens}. Therefore, we conclude that the proposed model has a good overall performance considering that it only applies acoustic data from mosquito wingbeats for classification. We highlight that the accuracy of the proposed multiclass model exceeds the one obtained in \cite{mukundarajan2017using}, which was 65\% using geographic metadata. More importantly, our results are close to the 80\% accuracy achieved by \cite{chen2014flying}, who considered only ten species of mosquitoes and used geographic metadata, whereas our approach is based solely on wingbeat sounds.

\subsection{Performance analysis of the ensemble classifier}
\label{subsec:performance-ensemble}

\begin{figure}[ht]
    \centering
\subfigure[]{%
    \centering
    \label{fig:eval:ensemble3:accuracy}%
    \includegraphics[width=0.42\columnwidth]{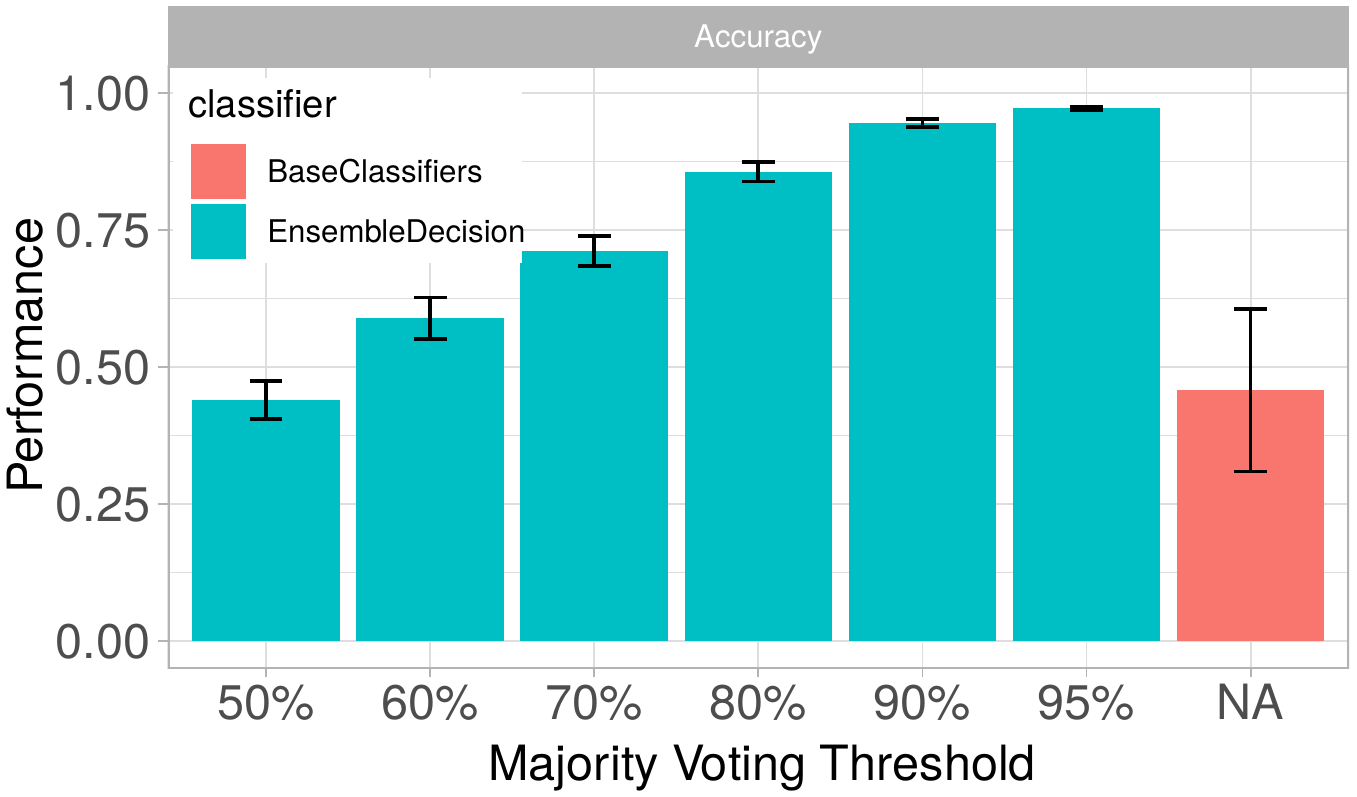}
}%
\quad
\subfigure[]{%
    \centering
    \label{fig:eval:ensemble3:f1measure}%
    \includegraphics[width=0.42\columnwidth]{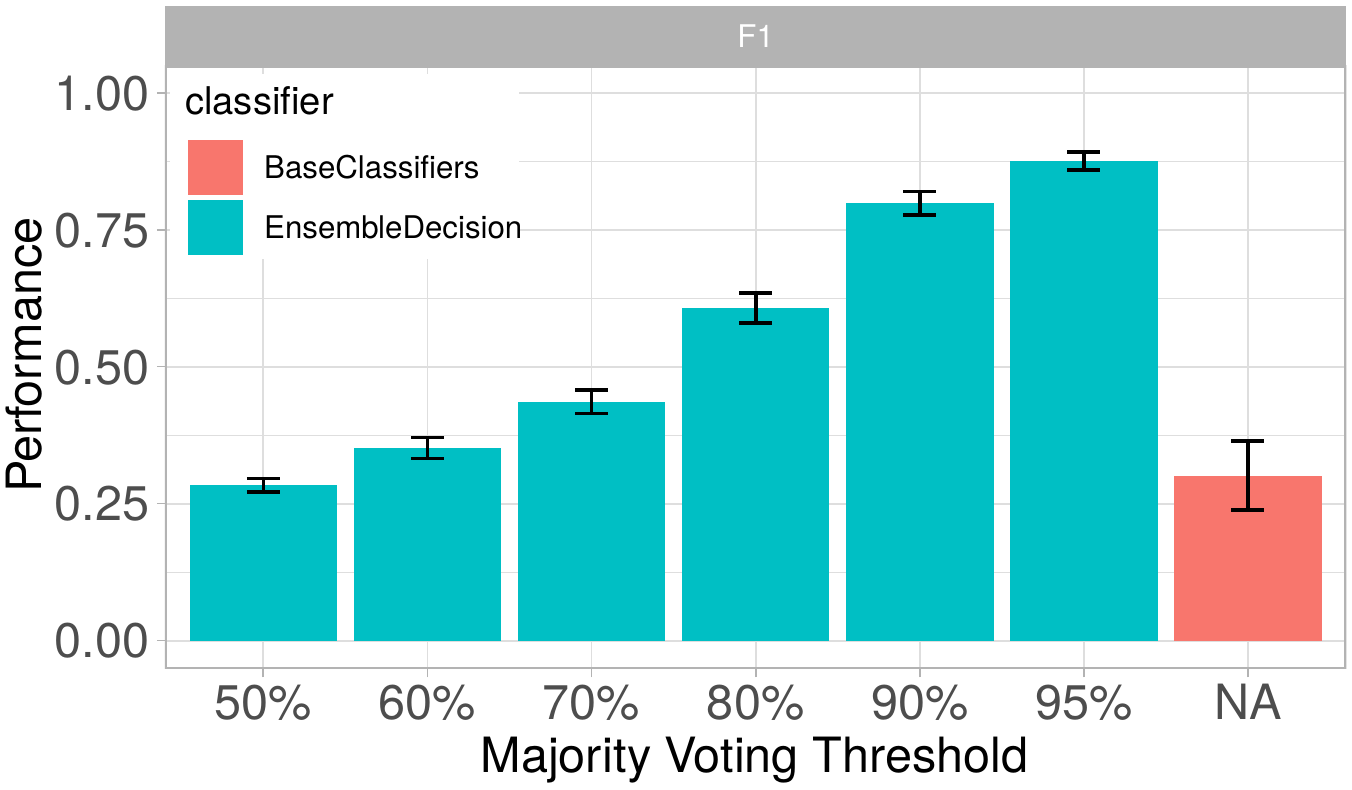}
}%
\quad
\subfigure[]{%
    \centering
    \label{fig:eval:ensemble3:precision}%
    \includegraphics[width=0.42\columnwidth]{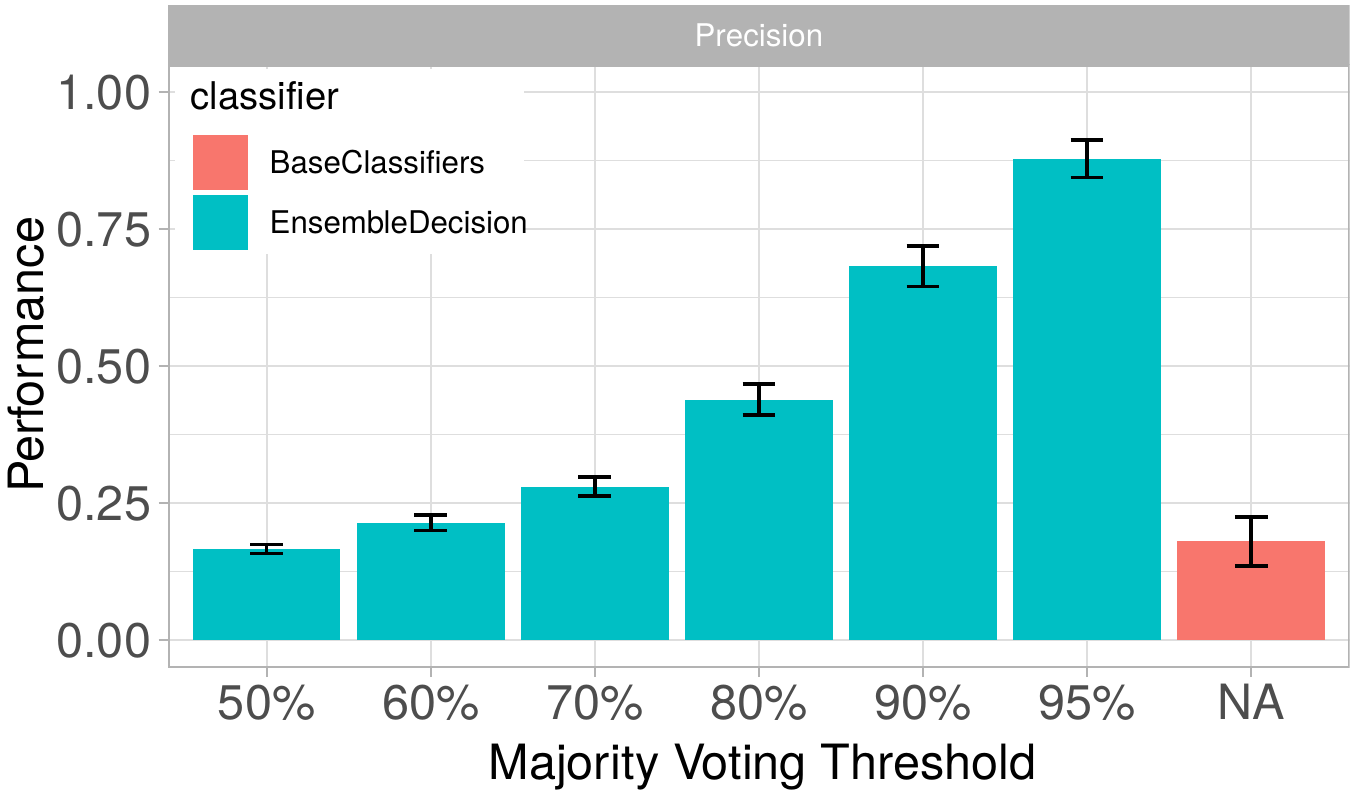}
}%
\quad
\subfigure[]{%
    \centering
    \label{fig:eval:ensemble3:recall}%
    \includegraphics[width=0.42\columnwidth]{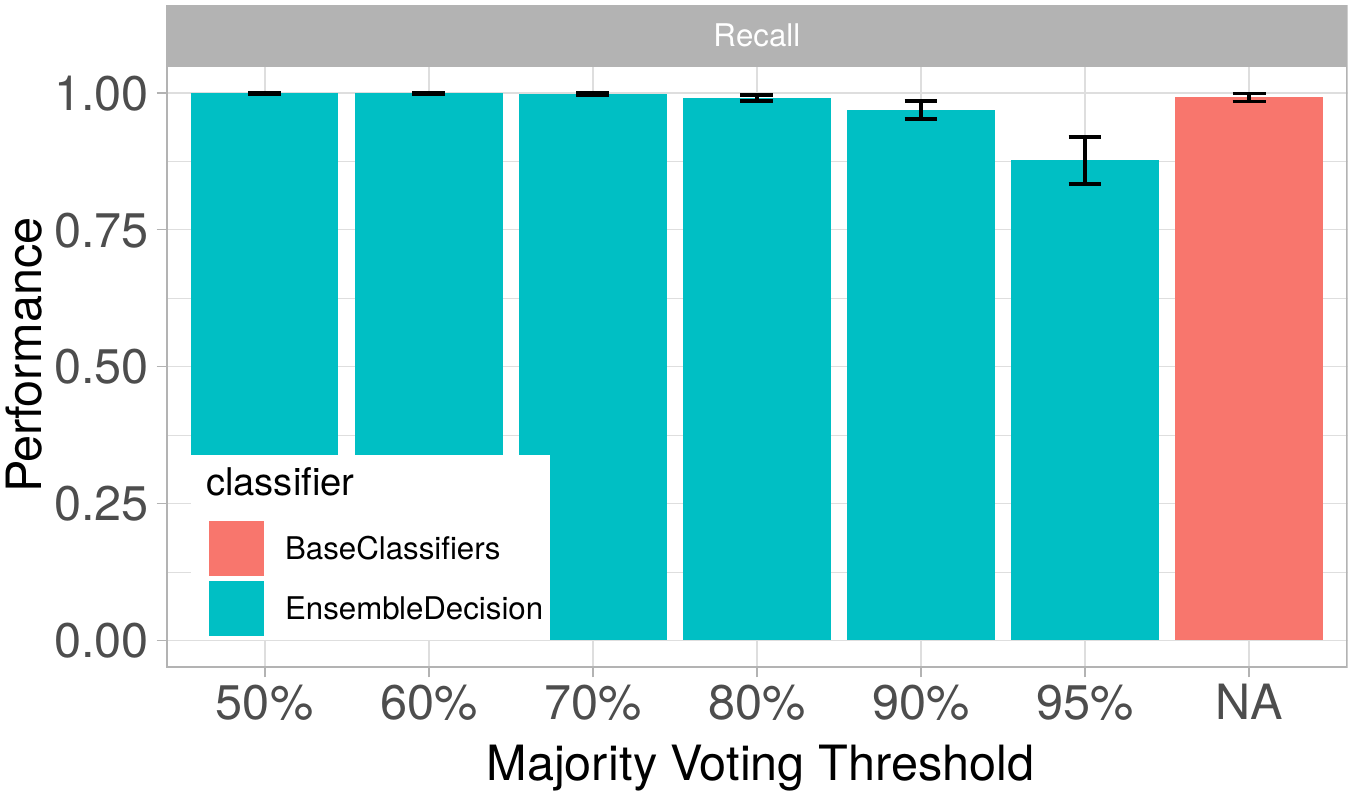}
}%

\caption{Mean and standard deviation performance of the ensemble model according to accuracy, F1-measure, precision, an recall, for different thresholds of majority voting. As baseline, we use the average performance across base classifiers (salmon bar), for which voting threshold was not applicable (NA).}
     \label{fig:eval:ensembleThresholds}
\end{figure}

Our last approach to identify \aedes from wingbeat recordings trains a set of 22 binary classifiers using model configuration \#8 and the CNN architecture described for the binary classifier (see Section~\ref{sec:approach:ensemble}). Each classifier is trained with \aedes as the positive class, alternating the negative class among all available species. In what follows, we refer to the negative class in a general way as non-\aedes. The ensemble classifier is based on a majority voting with a pre-defined threshold specifying the minimum percentage of votes \aedes must have to be considered as the ensemble's output (Figure~\ref{fig:workflow}). In our experiments, the threshold was varied between 50\% (simple majority) to 95\%. The motivation to evaluate a threshold that nears the complete agreement among base classifiers is the common overlap among species' frequency distributions, as demonstrated by \cite{mukundarajan2017using}, which may demand more strict thresholds to avoid misclassifications.

Results for this analysis are shown in Figure~\ref{fig:eval:ensembleThresholds}. In general, base classifiers (salmon bars) have a poor performance according to the distinct metrics. Except for the recall (Figure~\ref{fig:eval:ensembleThresholds}-d), which is close to 100\% with good stability, models are inefficient in terms of accuracy and F1-measure, mainly because of the extremely low values for precision (Figure~\ref{fig:eval:ensembleThresholds}-c).  
Unlike the previous binary approach, in which all the non-\aedes species are represented in the negative class, in this approach, each classifier tries to identify \aedes sounds based on the comparison against a single, specific non-\aedes species. The source of confusion arises when trying to classify new instances that belong to a class different from both the positive and negative classes used for model training: the wingbeat sound for these species may turn out to be more similar to the \aedes pattern than to the specific non-\aedes species used for that particular model, resulting in many false positive predictions. This may be caused by the large and frequent overlap of \aedes's wingbeat frequency distribution in relation to other species \citep{mukundarajan2017using}. Moreover, it should be pointed out that when considering non-\aedes classes individually (and not jointly, as in the binary classifier), the imbalance between classes favors \aedes, which becomes the majority class in all the 22 trained models and may introduce a bias to classify new data as specimens of this species.

In fact, this behavior may be perceived by the performance obtained when varying the majority voting threshold from a simple to a strict majority vote (\ie threshold of 95\%). For a simple majority, there is no performance gain in contrast to base classifiers. The only advantage observed may be described in terms of lower variance as compared to  base classifiers, which is an expected feature of ensemble methods. It is reasonable to assume that if 50\% of base classifiers predict a new instance as belonging to the \aedes class, it may still refer to a false positive if the species representing the true label of this instance has, in general, a wingbeat frequency distribution closer to \aedes than to the non-\aedes classes. 
Interestingly, the ensemble classifier results clearly improved with the increase of the threshold. From the threshold of 60\%, the ensemble classifier already surpasses the average performance of base classifiers for all metrics, although results are not as satisfactory as those obtained by our previous approaches. 

When considering the trade-off between precision and recall, we defined as the best ensemble model the classifier trained using the 90\% threshold for the majority voting, \ie requiring 20 out of the 22 votes in \aedes to classify instances in the positive class. This ensemble classifier is able to recover 96.82\% ($\pm$1.62) of positive instances in the testing set with a precision of 68.19\% ($\pm$3.67) and an overall accuracy of 94.56\% ($\pm$0.77).
Although the 95\% threshold increases the precision to 87.88\%, it reduces the recall (\ie the sensitivity for detection of \aedes) to 87.62\%, which is our primary interest in the current work.

\begin{figure}[h]
    \centering
\subfigure[]{%
    \centering
    \label{fig:eval:ensemble2:acc-f1}%
    \includegraphics[width=1\textwidth]{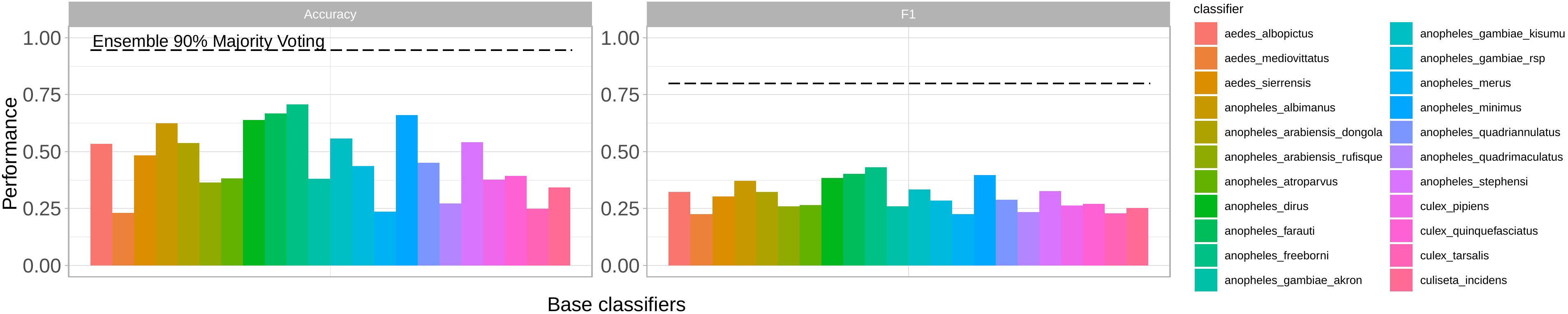}
}%
\quad
\subfigure[]{%
    \centering
    \label{fig:eval:ensemble2:prec-recall}%
    \includegraphics[width=1\textwidth]{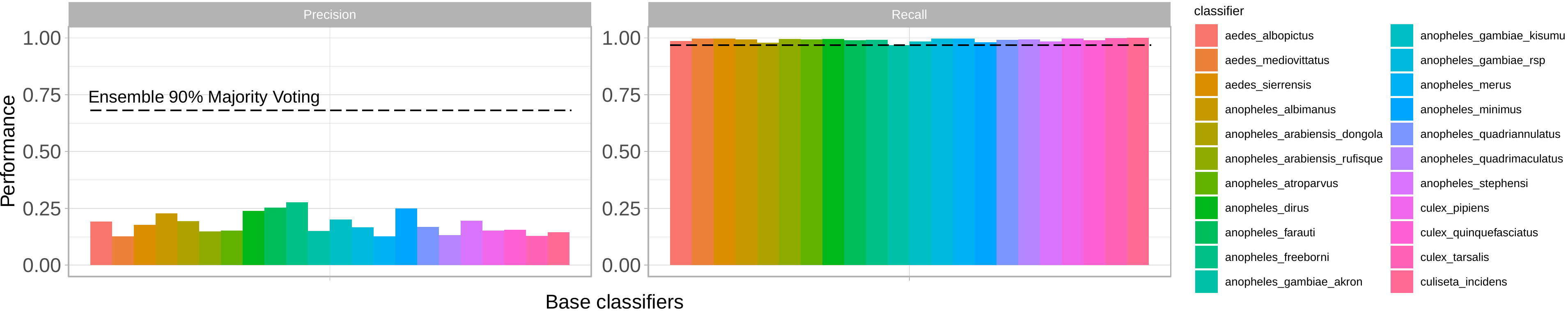}
}%
    \caption{Performance of the ensemble model with 90\% majority voting threshold (black dashed line) as compared to the base classifiers (colored bars). Each base classifier is identified by the non-\aedes species used as negative class. Values represent the average computed from 10 repetitions of training and testing the models.}
     \label{fig:eval:ensembleBaseClass}
\end{figure}

In Figure~\ref{fig:eval:ensembleBaseClass} we show the performance of the ensemble classifier with 90\% majority threshold (black dashed line) in contrast to the individual evaluation of the 22 base classifiers. Values represent the mean across the 10 repetitions of training and testing. Whereas the ensemble's recall is comparable to those obtained by base classifiers, there is a huge gain in performance in terms of accuracy, precision, and F1-measure. Accuracy for base classifiers varied between 23.05\% (\emph{Aedes mediovittatus}) and 70.67\% (\emph{Anopheles freeborni}), which is much lower than the 94.56\% achieved by this ensemble. Even more noticeable, the ensemble strategy had a substantial impact over precision: the increase in this metric was up to 5.39 times in relation to the \emph{Aedes mediovittatus}, which presented the lowest precision (12.64\%), and 2.46 times compared to \emph{Anopheles freeborni}, which showed the highest precision (27.65\%). 

We note that precision is low for base classifiers because, proportionally, most of the instances in the testing set do not belong to any of the two classes used for training a specific binary classifier. This is a more challenging scenario for classification, given that testing data very often comes from a distribution that differs from the training data. When predicting the label for these instances that do not follow the same distribution, the model will approximate their pattern to of one of the two captured in the training phase (\ie either \aedes or non-\aedes), and occasionally will (wrongly) indicate \aedes as the most similar pattern. To illustrate this situation, \emph{Culiseta incidens} may be taken as an example, given that it has the lowest frequency interval among all species (\ie 200 - 370Hz) \citep{mukundarajan2017using}. When an instance of this class is analyzed by the base classifier trained with \aedes and \emph{Anopheles minimus} data, for example, it may be predicted as \aedes given that their distributions are closer if we compare only the sound captured from the wingbeats. 
Thus, by raising the voting threshold, we impose to the ensemble model that to be classified as \aedes, the pattern from the input test instance has to be very close to the \aedes and different from most of the other species that we have as a basis for comparison, reducing the sources of uncertainty.

\section{Final Considerations}
\label{sec:conclusions}

Controlling the incidence of mosquito-borne diseases in under-developed regions requires inexpensive and easy-to-use strategies that can raise awareness on the occurrence of such mosquitos in these regions. In this work, we explored the possibility of using machine learning techniques to detect \aedes, a mosquito responsible for transmitting several diseases that still plague poor communities. Our approach is based on the analysis of audio data captured from mosquitoes wingbeats and three distinct formulations for the model using CNNs: a single binary classifier (\ie \aedes vs non-\aedes), a 23-class multiclass classifier, and an ensemble classifier that combines the output from 22 base binary classifiers.

\begin{table}[t]
    \centering
    \resizebox{\textwidth}{!}{  
    \begin{tabular}{c|c|c|c|c|c|c|c|c} \hline \hline
\multirow{2}{*}{Model} & \multicolumn{2}{c|}{Accuracy (\%)} & \multicolumn{2}{c|}{Precision (\%)} & \multicolumn{2}{c|}{Recall (\%)} & \multicolumn{2}{c}{F1-measure (\%)} \\ \cline{2-9}
              & Avg. & Std. Dev. & Avg. & Std. Dev. & Avg. & Std. Dev. & Avg. & Std. Dev. \\ \hline \hline
Binary  & 97.65 & 0.55     & 90.63 & 4.33     & 88.49 & 6.68     & 89.29 & 2.91 \\
Multiclass - \aedes only  & \multirow{2}{*}{78.12} & \multirow{2}{*}{2.09} & 89.52 & 3.69     & 90.23 & 3.83    & 89.76 & 1.82\\
Multiclass - all mosquitoes  &  &   & 79.41 & 13.00     & 76.22 & 16.49    & 76.71 & 13.24\\
Ensemble (90\% voting threshold) & 94.56 & 0.77     & 68.19 & 3.67     & 96.82 & 1.62     & 79.95 & 2.13 \\
     \hline \hline
    \end{tabular}
    }
    \caption{Summary of performances obtained by the proposed strategies in the classification of \aedes based on its wingbeat sounds. For the multiclass approach, we show the results for the general model (\ie all mosquito species) and for \aedes class.}
    \label{tab:eval:summary}
\end{table}

The results achieved are promising, as summarized in Table~\ref{tab:eval:summary}. An accuracy above 94.5\% was achieved with the binary classifier and the ensemble classifier using the 90\% threshold voting. The best recall, which represents the sensitivity for detecting \aedes, was obtained with the ensemble classifier, although at the cost of a lower precision. The most balanced performance in the classification of \aedes in terms of precision and recall was observed in the binary and multiclass approaches, with F1-measure close to 90\%. We note that our multiclass classifier presented an accuracy of 78.12\% ± 2.09, comparable to prior work in which geographic metadata was used in addition to audio data for mosquito identification. Our results could be improved and probably surpass previous approaches, if metadata such as time and location of recording were incorporated into the model to mitigate classification errors due to high overlap in wingbeat frequency distributions among species.

In spite of the promising results, much work remains towards the development of an easy-to-use solution that can be widely used for crowd-sourcing the mapping of mosquito occurrence. One challenge is running a simplified but effective version of the model that could operate offline in a smartphone, without posing much overhead in terms of required processing power and battery usage. Another challenge is devising a user interface that enables the local population to quickly raise awareness on regions in which there is evidence of the circulation of \aedes. Moreover, the ML models themselves could be improved in some directions, such as using metadata as additional evidence in training the models (e.g., geolocation data), adopt strategies to adjust class imbalance, and explore different strategies for creating the ensemble classifier (\eg, weighted majority voting, stacking, among others).

\section*{Acknowledgements}
This work has been supported in part by funding from Conselho Nacional de Desenvolvimento Cient\'{i}fico e Tecnol\'{o}gico (CNPq) and Coordena\c{c}\~{a}o de Aperfei\c{c}oamento de Pessoal de N\'{i}vel Superior (CAPES) - Finance Code 001.

\bibliographystyle{cas-model2-names}

\bibliography{paper}

\end{document}